\documentclass[sigconf]{acmart}

\usepackage{booktabs} % For formal tables
\usepackage{algorithm}
\usepackage[noend]{algpseudocode}
\usepackage{graphicx}
\usepackage{subfigure}
\usepackage{url}
\usepackage{color}
\usepackage{multirow}
\usepackage{amsmath,amsfonts}
\usepackage{gensymb}
\usepackage{comment}
\usepackage{xspace}
\usepackage{balance}
\usepackage{lipsum}
\usepackage{fancyhdr}

\usepackage{xspace}

\newcommand{\note}[1]{{\color{black}{#1}}}

\newcommand{\squeezeup}{\vspace{-7pt}}

\acmYear{2020}

\begin{document}

%\date{}

%make title bold and 14 pt font (Latex default is non-bold, 16 pt)
%\title{\Large \bf Exploiting \Capname Capacity for Server Sprinting in Multi-Tenant Data Centers}

\title{A Note on Latency Variability of Deep Neural Networks for Mobile Inference}

 \author{Luting Yang}
 %\email{fyang050@ucr.edu}
 \affiliation{%
   \institution{UC Riverside}
 }

 \author{Bingqian Lu}
 %\email{webmaster@marysville-ohio.com}
 \affiliation{%
   \institution{UC Riverside}
 }

 \author{Shaolei Ren}
 %\email{webmaster@marysville-ohio.com}
 \affiliation{%
   \institution{UC Riverside}
 }

\maketitle

% Use the following at camera-ready time to suppress page numbers.
% Comment it out when you first submit the paper for review.
%\thispagestyle{empty}

\subsection*{Abstract}

Running deep neural network (DNN) inference on mobile devices,
i.e., mobile inference, has become a growing trend, making
inference less dependent on network connections and
keeping private
data locally.
The prior studies on optimizing DNNs
for mobile inference typically focus on the metric
of average inference latency,
thus implicitly assuming that  mobile inference  exhibits
little latency variability.
In this note,
we conduct a preliminary measurement study
on the latency variability of DNNs for mobile inference.
We show that the inference latency variability can become quite
significant in the presence of CPU resource contention.
More interestingly, unlike the common belief
that the relative performance superiority of DNNs on one
device
can carry over to another device and/or another level of resource contention,
we highlight that
a DNN model
 with a better latency performance than another model can become
outperformed by the other model when %the level of
resource contention
be more severe or running on another device.
%Our results imply that 
Thus, when optimizing DNN models for mobile inference, only
measuring the average latency may not be adequate;
instead, latency variability under various conditions should be accounted for, including but not limited to
different devices and different levels of CPU resource contention considered in this note.

\section{Introduction}

The unmatchable predictive power of deep neural networks (DNNs)
has been successfully attested by numerous applications, including
speech/image recognition, malware detection, health care, among others \cite{DNN_Book_Goodfellow-et-al-2016}.
Traditionally, due to the prohibitively large DNN model size
and computational requirement,
the inference tasks of DNNs for
resource-constrained mobile devices
are usually offloaded to data centers or distributed high-performance
servers. For example, a mobile device that needs to perform
inference (e.g., image style transfer) can
send its request to a remote cloud where pre-trained DNNs are hosted, and then subsequently receives the inference result
via communication networks.
While data centers still remain the mainstream platform
for DNN training, executing DNN inference entirely and directly
 on mobile devices (a.k.a., \emph{mobile inference} or on-device inference) has been quickly trending, as evidenced by the Facebook app that
integrates built-in DNNs (e.g., for real-time image
style transfer) and supports
mobile inference for billions of active users \cite{DNN_Facebook_Inference_HPCA_2019}.
Compared to offloading-based inference,
the key advantages of mobile inference include
being less reliant on network connection
and also, more importantly, better protecting privacy
by keeping user's (possibly sensitive) data locally
without transferring it to a remote cloud or platform.
%Thus, mobile inference is favorably being integrated with
% many apps,

\begin{table*}[!t]\small
	\centering
	\caption{Device Configuration}
	\label{table: device_details}
		\begin{tabular}{|c|c|c|c|c|c|c|c|c|}
			\hline
			\textbf{Device Name} & \textbf{CPU (GHz)} & \textbf{Cores} & \textbf{RAM (GB)} & \textbf{RAM Freq. (MHz)} & \textbf{OS} & \textbf{Display} & \textbf{Battery (mAh)}  \\
			\hline
			Samsung Tab A & 2  & 4 &   2 & 933 & 9.0 & 1280 x 800 & 5100 \\
			\hline
			Samsung Tab S5e & 2 & 8 &  4  & 1866 & 9.0 & 2560 x 1600 & 7040 \\
			\hline
			%Lenovo Moto Tab & 2 & 8 & 1 & & 2 & 933 & 7.0 & 1200 x 1920 & 7000 \\
			%\hline
			
		\end{tabular}
\end{table*}

\begin{table*}[!t]\footnotesize

\centering

%\sffamily

\caption{DNN Model Configuration with Estimated Parameters}

\label{table:model_details}

\begin{tabular}{|l|c|c|c|c|c|c|c|}

\hline

\textbf{Model Name} & \textbf{Million MACs} & \textbf{Million Param.} & \textbf{Size (Mb)} & \textbf{Accuracy} &  \textbf{Nodes} & \textbf{Layers} \\

\hline

MobileNet\_V1\_0.75\_192\_Quant (\textbf{V1Q}) & 233 & 2.59 & 2.6 & 66.1\%  &  5984  & 30 \\

		\hline

MobileNet\_V2\_1.0\_224\_Quant (\textbf{V2Q}) & 300 & 3.47 & 3.4 & 70.8\%  &  2810  & 25 \\

\hline

MobileNet\_V1\_0.75\_192 (\textbf{V1F}) & 233 & 2.59 & 10.3 & 67.1\%  &  5984  & 30 \\

\hline

MobileNet\_V2\_1.0\_224 (\textbf{V2F}) & 300 & 3.47 & 14 & 90.6\%  &  2810  & 25 \\

\hline

Inception\_V3\_Quant (\textbf{V3Q}) & 5900 & 23.9 & 23 & 77.5\%  &  18400  & 159 \\

\hline

Inception\_V4\_Quant (\textbf{V4Q}) & 16800 & 55.8 & 41 & 79.5\%  &  32480  & 160 \\

\hline

Inception\_V3  (\textbf{V3F}) & 5900 & 23.9 & 95.3 & 77.9\%  &  18400  & 159 \\

\hline

Inception\_V4  (\textbf{V4F}) & 16800 & 55.8 & 170.7 & 80.1\%  &  32480  & 160 \\

\hline

%NASNet large & 23800 & 27.6 & 355.3 & 82.6\% & 1170 & 46704  & 50 \\

%	\hline

\end{tabular}

\end{table*}

The emergence of DNN-powered mobile inference is made possible by
the increasingly more powerful
computing
capabilities of mobile systems as well as  recent progress in
DNN model compression
\cite{DNN_Compression_CharacterizingDNN_Mobile_Inference_TianGuo_arXiv_2019,DNN_Compression_PatDNN_Mobile_YanzhiWang_ASPLOS_2020,DNN_Compression_SongHan_ICLR_2016,DNN_EdgeInference_IntermittentEmbeddedSystem_CMU_ASPLOS_2019_Gobieski:2019:IBE:3297858.3304011}.
Concretely, state-of-the-art DNN neural architecture
search and model compression
techniques, such as
weight pruning/quantization in either structure
or non-structured manners \cite{DNN_Compression_Structured_YiranChen_NIPS_2016_10.5555/3157096.3157329,DNN_FBNet_HardwareAwareConvNetDesign_CVPR_2019_Wu2018FBNetHE,DNN_Compression_AutoCompress_YanzhiWang_AAAI_2020,DNN_Compression_SongHan_ICLR_2016,DNN_Compression_PCONV_Sparsity_YanzhiWang_AAAI_2020,DNN_EdgeInference_YiyuShi_NatureEle_2018_xu2018scaling},
can remarkably reduce the DNN size %by a factor of 10+
and cut
the \emph{average} inference latency to the order
of tens/hundreds of milliseconds or even less,
yet  without significantly
compromising the inference accuracy. To improve performance, an additional term accounting for the latency is often factored
 into the loss function or constraints during DNN training/compression \cite{DNN_FBNet_HardwareAwareConvNetDesign_CVPR_2019_Wu2018FBNetHE,DNN_ModelCompressionAcc_Survey_IEEE_TSP_2018_8253600}.

 %
%typically
%achieved by accounting for

While the average latency of DNN inference
is an important performance metric,
 latency variability is also equally, if
not more, crucial for users' quality of experience.
Consider a simple scenario where a mobile user
live streams its activities with DNN-based
style transfer in real time. Clearly,
a low average latency but a high variability in
DNN inference can easily make the user feel frustrated with the app.
%In this paper, we take the position that
%Therefore, we take the position that in addition
%to the inference accuracy and average latency, the inference latency variability
%of DNNs on mobile devices must also be carefully addressed.
%Our position is also reinforced by a recent Facebook study
% which presents an open challenge of significant performance variability \cite{DNN_Facebook_Inference_HPCA_2019}.
% that the inference latency of the same DNN model
%but on different

 In practice, it is rare to have a real mobile app, except
for toy projects, which does nothing but only DNN inference
(e.g., pure image classification without other functionalities). Instead,
DNN inference is typically combined with
other tasks: DNN-based vision and tracking
are only part of mobile augment reality applications \cite{DNN_AugmentReality_JiasiChen_SenSys_2019_10.1145/3356250.3360044},
whereas DNN-based style transfer is running concurrently with
communications activities for live streaming applications.
More generally, DNN inference on mobile devices can be executed under
an extremely diverse set of runtime conditions,
such as time-varying resource contention caused by
concurrent threads, different numbers of background services,
dynamic system settings   (e.g., CPU
speed and battery status) and even ambient temperature,
which are all potentially interfering factors for DNN inference and can  contribute to the inference latency variability.
 %different number of background apps, the degree
%of resource contention, system setting (e.g., CPU
%speed), battery status, and even ambient temperature.

Nonetheless, the existing studies on DNN model compression
and neural architecture search for mobile inference
\cite{DNN_Compression_Structured_YiranChen_NIPS_2016_10.5555/3157096.3157329,DNN_FBNet_HardwareAwareConvNetDesign_CVPR_2019_Wu2018FBNetHE,DNN_Compression_AutoCompress_YanzhiWang_AAAI_2020,DNN_Compression_SongHan_ICLR_2016,DNN_Compression_PCONV_Sparsity_YanzhiWang_AAAI_2020,DNN_EdgeInference_YiyuShi_NatureEle_2018_xu2018scaling}
typically focus on and measure the average inference latency
in a static (and often idealized) environment where the
DNN model is running with little interference from the aforementioned factors.
For example, %like in many research studies,
the reported inference latency
of mobile DNN models hosted on TensorFlow \cite{DNN_TensorFlowLite}
only mentions a single performance value ``measured on Pixel 3 on Android 10'' without
further details regarding the actual runtime condition.
%Thus, little is known about the actual inference latency of the
%DNN models
Thus, the average inference latency measured in
an idealized setting can only represent
the best-case performance, and fails
to capture the actual latency in a practical environment
with significant variabilities.
%the actual
% latency can be far off the reported values when running
% DNN inference in a practical environment.

In this note, we conduct a preliminary measurement study
on the latency variability of DNNs for mobile inference.
In particular, we consider eight popular DNNs
and run them on two mobile devices (listed
in Tables~\ref{table:model_details} and~\ref{table: device_details})
under a diverse set of runtime conditions.
We explicitly focus on how the background apps and CPU
resource contention (created by concurrent threads within
the same app as DNN inference)
affect the inference latency of DNN-based image classification on mobile
devices. We find that the number
of background apps has little impact on the inference latency,
and the inference latency variability is reasonably small in
a \emph{static} environment with little CPU resource contention for most
DNN models under our investigation.
Nonetheless,
our measurement results also
highlight that when the number of concurrent threads varies and creates
different levels of CPU resource contention, both the average latency
and latency variability can vary significantly. Interestingly
and also counter-intuitively, one DNN model
 with a lower average latency
and/or latency variability than another model can become
outperformed by the other model when running
on another device and/or %the level of
CPU contention
becomes more severe.

Our study implies that only measuring
the average latency of a DNN model in
a static and contention-free environment
 is
inadequate to fully quantify the actual performance for mobile inference:
\textbf{the relatively better latency performance of a DNN model
than that of another model under one condition does not necessarily carry
over to another device and/or another level of CPU resource contention.}
%a lower average latency
%for mobile inference in an idealized condition
%does not necessarily translate into a lower average latency
%and/or latency variability in a more practical condition with
%time-varying resource contentions, even on the same device.
%To our knowledge,
%there has been little understanding about
%the latency variability under different resource contention levels
%%how the resource contention affects the DNN latency
%for DNN-based mobile
%inference.
This warrants more investigation into the urgent issue of latency  variability that is crucial for user experience.
Thus, we take the liberty that, in addition to the already-considered metrics such as
inference accuracy and average latency,
the latency variability under a diverse set of conditions
 with time-varying resource contention levels should also be measured and compared
 when optimizing DNNs  for mobile inference.
%By this paper, we call for more investigation into the urgent issue of latency
% variability that is crucial for user experience.

% while the vast majority
%of the existing studies on DNN model compression
%and neural architecture search for mobile inference
%measure the average latency in a static condition,

% having a lower average latency
%does not mean the DNN model will consistent

% There are many factors that can potentially affect the mobile inference latency, including the number of background apps, the degree
%of resource contention, system setting (e.g., CPU
%speed), battery status, and even ambient temperature.

\begin{figure*}[!t]
	\centering
	\subfigure[Inception V3F on Tab A]{\label{fig:A_Inception_V3_float_App_based}\includegraphics[trim=0 0 0 0, clip, width=0.24\textwidth,clip]{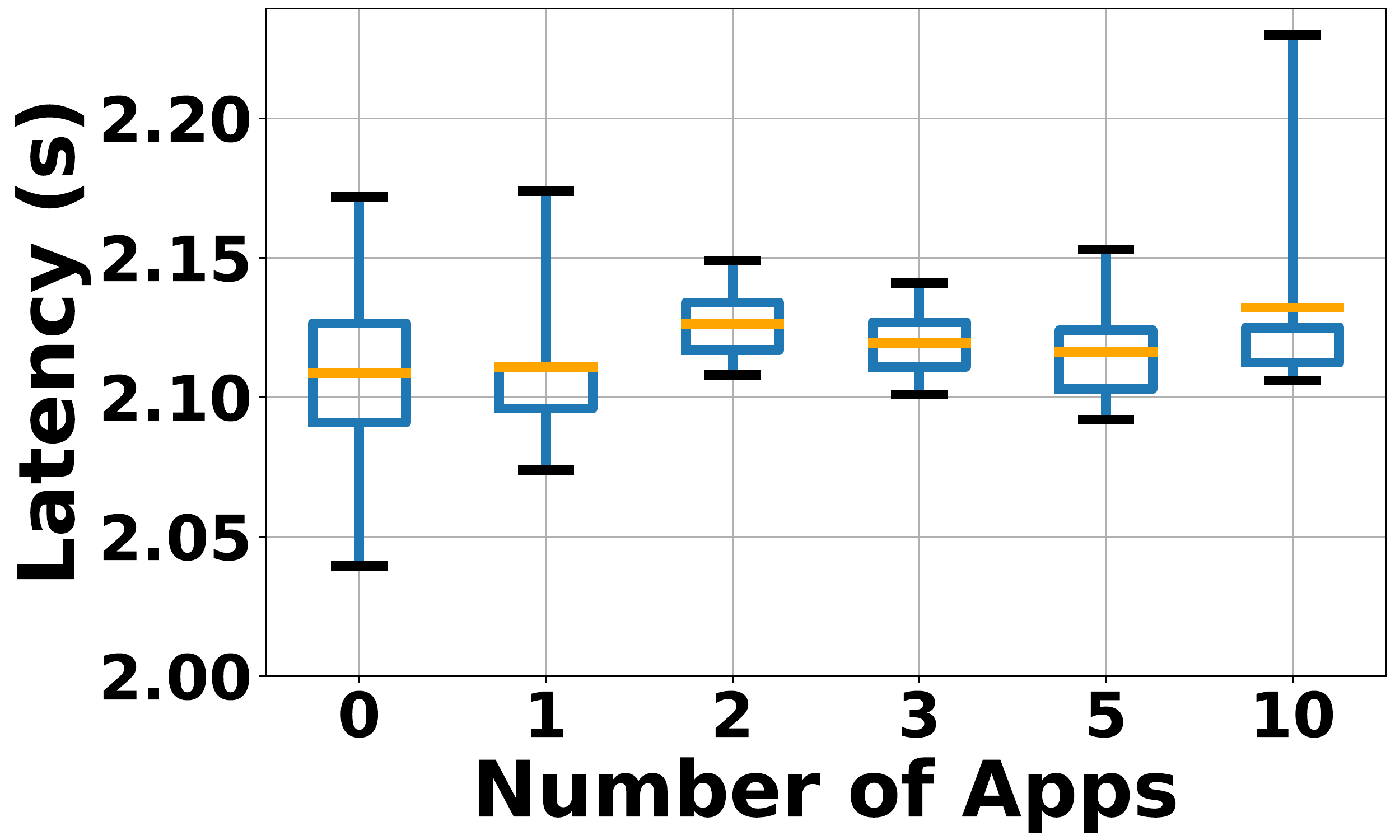}}
	\subfigure[Inception V3F on Tab S5e]{\label{fig:S5_Inception_V3_float_App_based}\includegraphics[trim=0 0 0 0, clip, width=0.24\textwidth,clip]{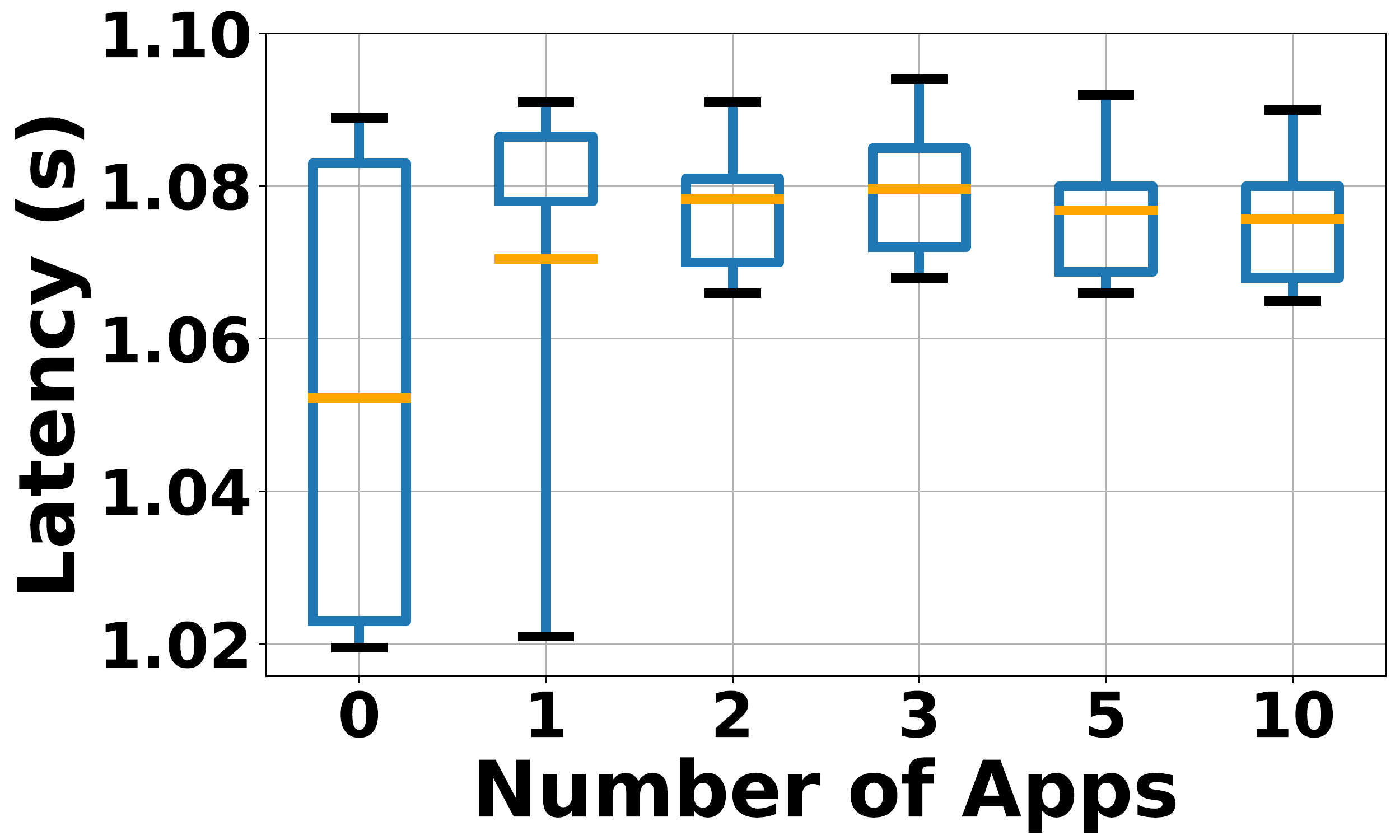}}
\subfigure[MobileNet \note{V2Q} on Tab A]{\label{fig:A_Inception_V3_float_App_based}\includegraphics[trim=0 0 0 0, clip, width=0.24\textwidth,clip]{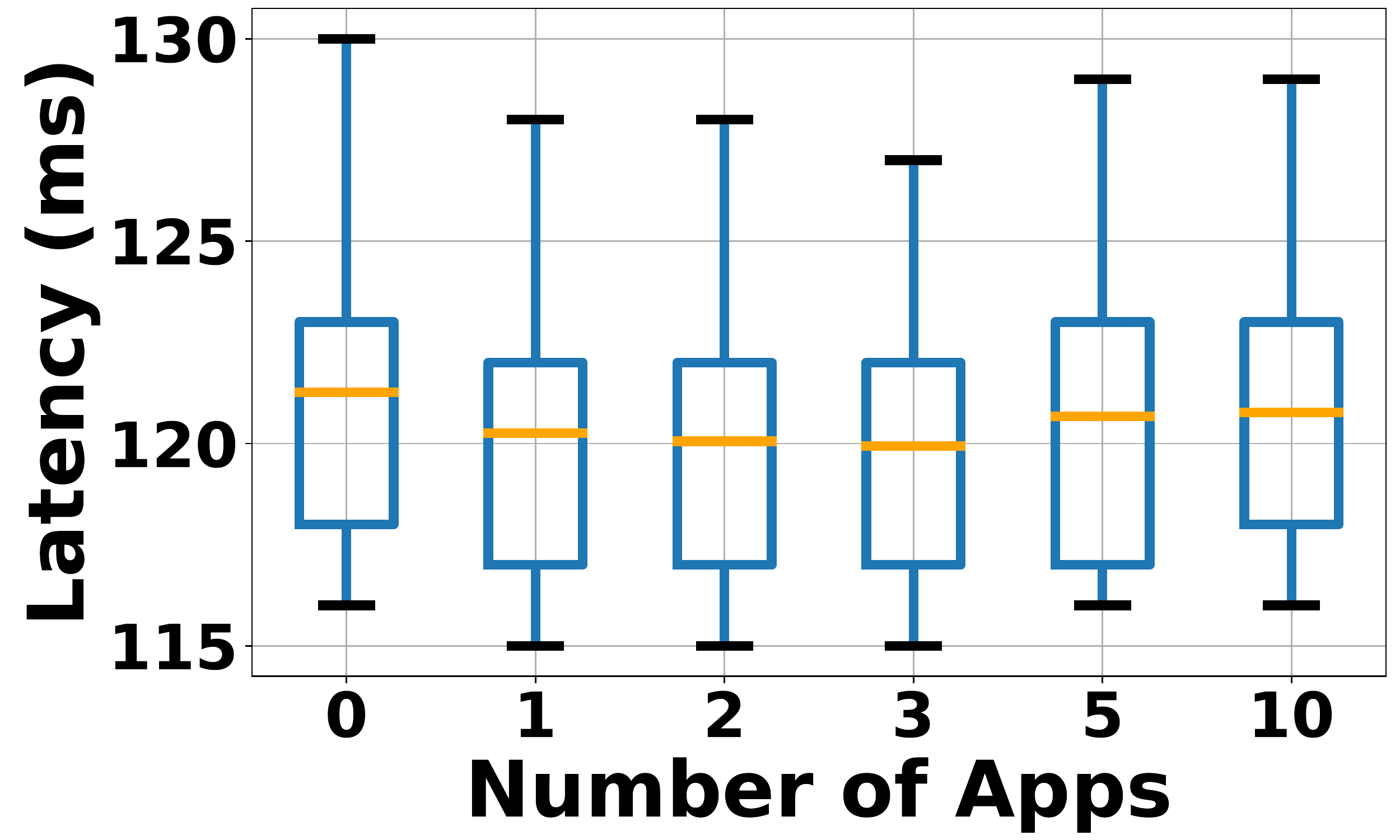}}
	\subfigure[MobileNet \note{V2Q} on Tab S5e]{\label{fig:S5_Inception_V3_float_App_based}\includegraphics[trim=0 0 0 0, clip, width=0.24\textwidth,clip]{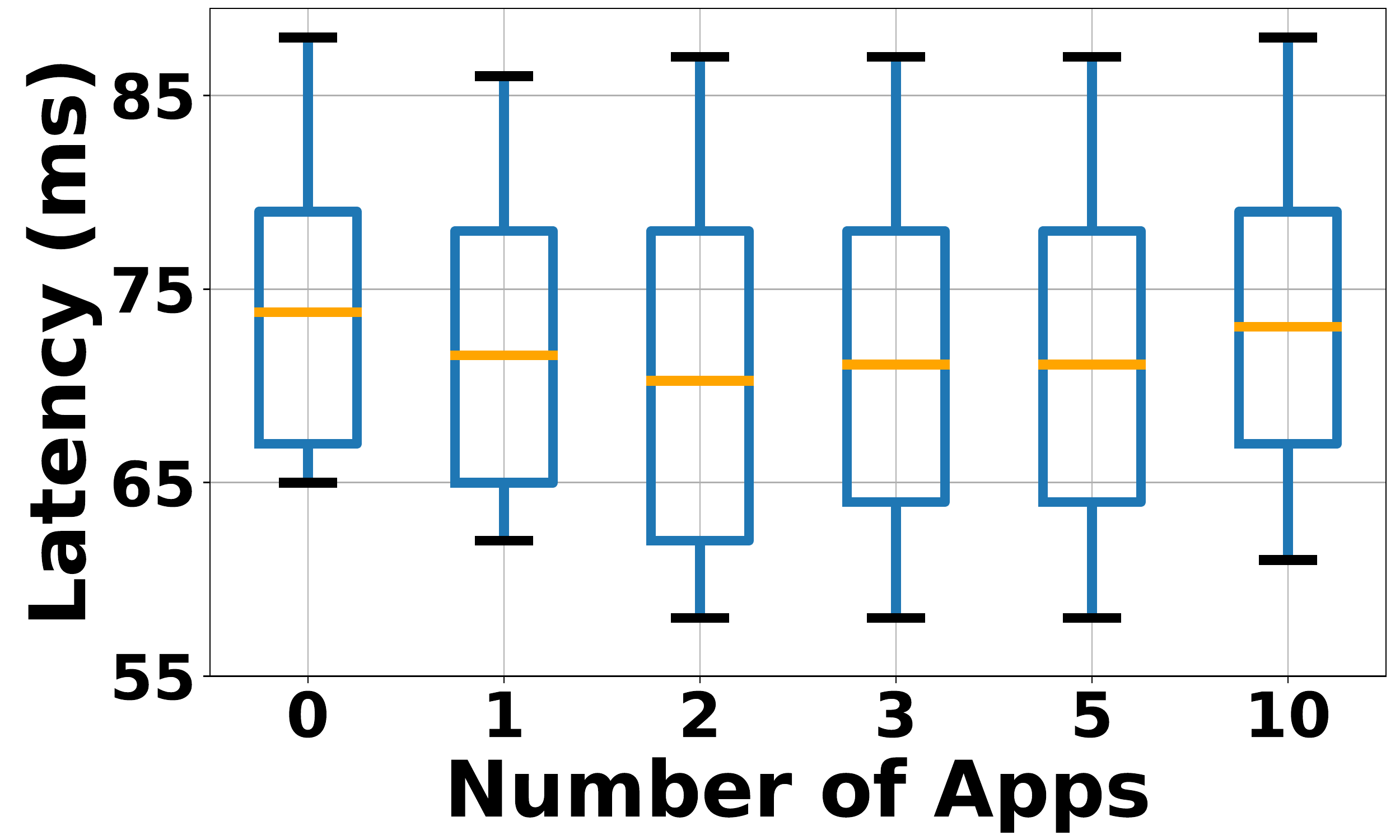}}
\squeezeup
	\caption{Inference latency  with different numbers of background apps.}\label{fig:background_app}
\end{figure*}

\begin{figure*}[!t]
	\centering
	\subfigure[CPU and memory usage]{\label{fig:S5_Mobilenet_V2_float_cpu_mem_trace}\includegraphics[trim=0 0 0 0, clip, width=0.48\textwidth,clip]{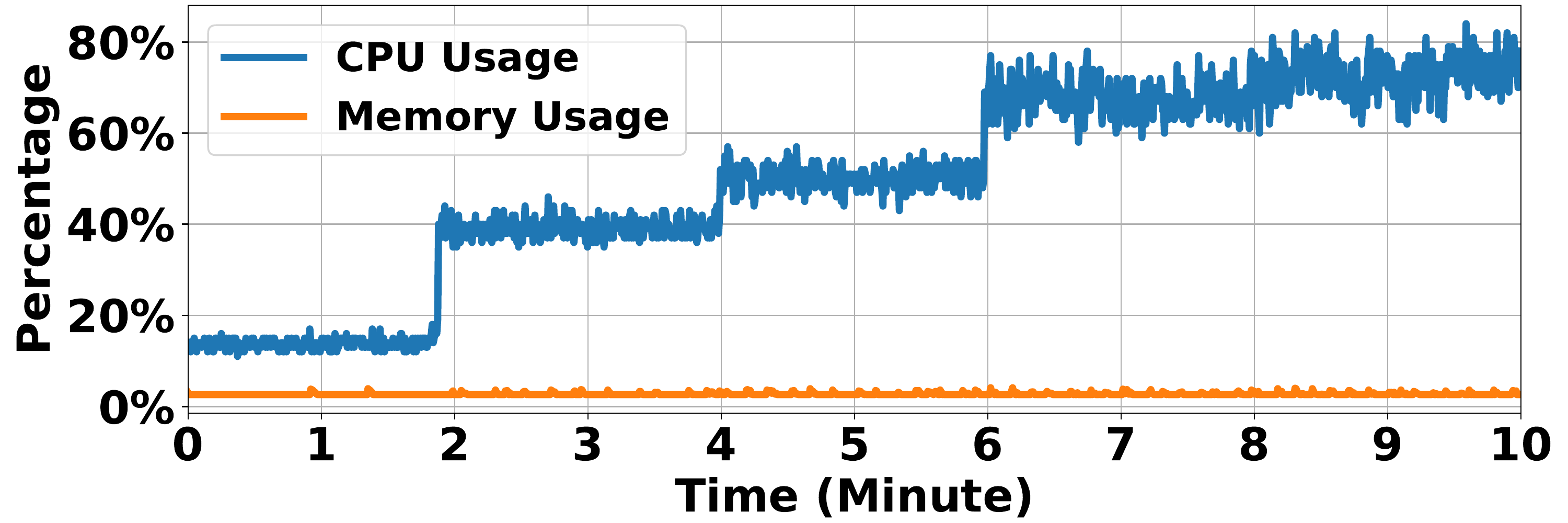}}
	\subfigure[Inference latency]{\label{fig:S5_Mobilenet_V2_float_latency_trace}\includegraphics[trim=0 0 0 0, clip, width=0.48\textwidth,clip]{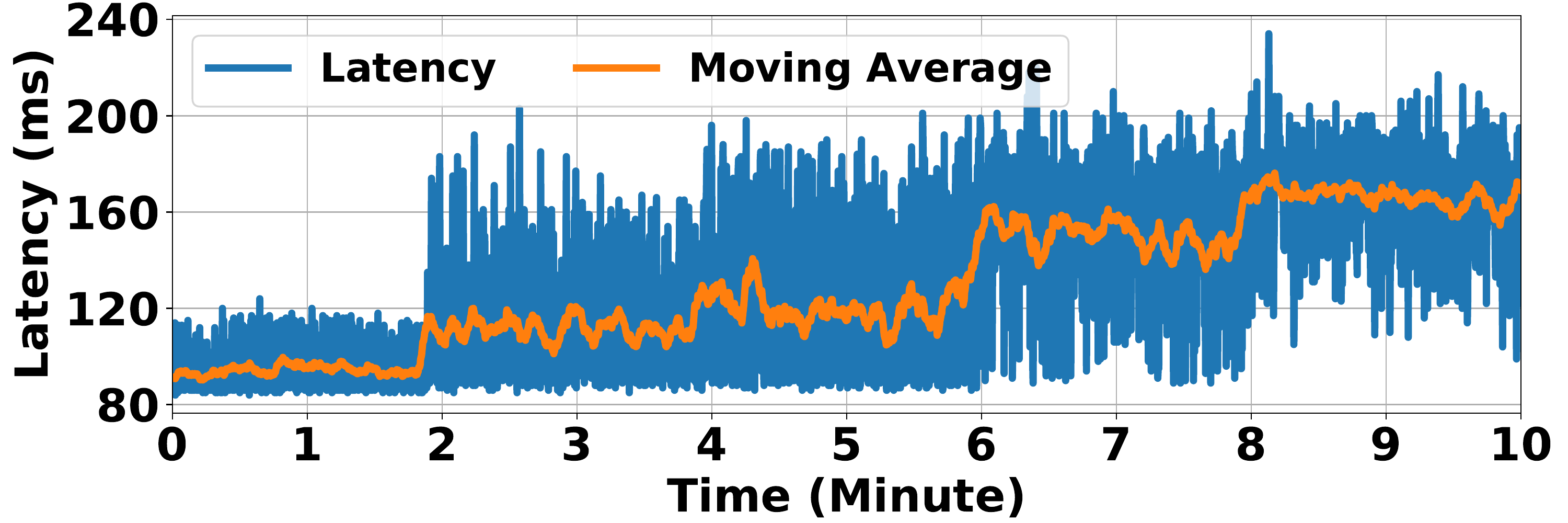}}
\squeezeup
	\caption{Traces of CPU/memory usage and latency of MobileNet V2F on Samsung Tab S5e.}\label{fig:thread_trace}
\end{figure*}

\section{Methodology}

%In this section, we provide some preliminaries on the latency
%variability of DNNs for mobile inference, followed by the methodology
%of our measurement.

%\subsection{Preliminaries on Latency Variability}

In general, the inference latency is jointly affected by
the DNN model, mobile device the DNN is running on, as
well as the device's runtime condition and resource management policies. For example,
mobile devices have very diverse configurations and thus
exhibit different inference latencies even for the same DNN model: high-end devices can have
powerful CPUs/GPUs along with
%dedicated graphic processing units (GPUs)
%and even
purpose-built accelerators %processing units (e.g., edge TensorFlow processing units)
to speed up inference,
whereas nearly 75\% of Facebook's mobile users
 are powered by CPUs of at least seven years old \cite{DNN_Facebook_Inference_HPCA_2019}.
Moreover, the inference latency can also %differ
%significantly, depending on
be subject to runtime system condition (e.g., number of
concurrent threads) and resource management policies.

%Last but not least, users have different preferences
%towards different metrics: some users are more energy-sensitive due to limited %battery capacities, whereas
%others like to trade energy consumption for latency.

Given a DNN model running on a mobile device, we focus on the impact of two runtime factors --- the number
of background apps and the level of CPU contention --- on the inference latency.
Our experiment setup is described as follows.

%\subsection{Methodology of Measurement}

\textbf{Overview.} We build an image classification app hosted on TensorFlow Lite for Android \cite{DNN_TensorFlowLite}, which continuously takes input images and provides
real-time classification results. The app is installed on two mobile
devices, whose details are listed in Table~\ref{table: device_details}.

In line with the official source
code, the inference latency is calculated as the sum of time to load the input image
 and the time to run model inference. For a model on a device, it takes
 a small (roughly constant) time to lead each image regardless
 of experiment conditions we have tested.
 For example, on Samsung S5e, the per-image loading times for
 MobileNet V2Q  and Inception V4Q are only about 8ms and 15ms,
 respectively. For each DNN model under each condition, we run more than 1,000
 inference tasks.
We log the inference latency for each image and save it  for offline analysis.

At the core of the app is a pre-trained DNN model.
In this note, we choose eight official models from TensorFlow Lite
in two categories: MobileNet models which are lightweight and specifically tailored
to resource-constrained mobile devices at the expense of inference accuracy,
and Inception models which reduce the computational cost while maintaining
a good accuracy performance \cite{DNN_TensorFlowLite}. The details
of the DNN models are listed in Table~\ref{table:model_details}.
Although we can choose any other DNN models including
more advanced ones for measurement, we focus on these eight official models
because they are popularly used as benchmarks.

Unless otherwise stated, all our latency measurement results will
be shown in error bars, indicating the 5th, 25th, 75th, and 95th percentile
as well as average latencies.

\textbf{Background apps.} To investigate the impact of background apps
on inference latency, before running our image classification app
in the foreground mode, we open and then turn into the background model
the following apps in order:
Facebook, Youtube, Messenger, Google Search, Google Maps, Instagram,
Snapchat, Google Play, Gmail, Pandora Radio,
which are top 10 most used apps in the U.S.
If we consider $n$ background apps, we will put the top $n$ apps into the background.

%Core of this application is the pre-trained DNN model deployed to recognize input images.
%A set of pre-trained models that excel in distinct metrics (e.g., inference latency, accuracy, model size) are also provided by Tensorflow Lite \cite{hosted_models}.
%This way, by deploying different DNN models in the applicaton, we can profile the model performance indirectly.

%We set up our experiment platform% utilizing the image classification application providede by Tensorflow Lite \cite{Tensorflow_Lite_Image_Classification} on Android system, which continuously takes input pictures and give real-time classification results, accuracy, and latency.

\textbf{CPU contention.} In most practical apps, DNN inference is only part
of a more complex function (e.g., style transfer for fun chat in Facebook),
with multiple concurrent threads that all dynamically access system resources at runtime.
Nonetheless, our simple image classification app is for research purpose
and hence it excludes all other functions. To
enable random CPU contention under a controlled setting, we launch multiple concurrent threads within our app.
Specifically, %each thread performs CPU-intensive computation (i.e.,
%random matrix multiplication in our code) in an on/off manner:
we divide the time into slots each having 5ms,
and for each time slot, every thread performs CPU-intensive computation (i.e.,
random matrix multiplication in our code)  with a probability
of $x\%$ and stays idle with a probability of $1-x\%$.
All the threads are run concurrently along with the
DNN-based image classification in the foreground mode
within the same  app.

%Then we further test if activity within the same application would cause model performance variability, which is implemented by adding threads running resource-consuming tasks to the image classification demo.
%This experiment is critical because practical machine learning applicaions commonly run multiple different types of tasks together with DNN model inference, while what is provided by Tensorflow Lite is a single-function one.
%Concretely, we create two kinds of threads which perform random matrix multiplication and network packet transmission with Google.com respectively.
%We also assign a utilization rate to each thread to control the thread busy and idle time.

%Note that aligning with what the official source code \cite{image_Classification_scource_code} defines, we treat the confidence score as classification accuracy, and the inference latency is calculated as the sum of time cost to load the input image and time to run model inference.
%Core of this application is the pre-trained DNN model deployed to recognize input images.

%Next we set up two types of experiments to test the model robustness under practical user usage cases.
%We firstly run the above image classification application simultaneously with various background activity within the system, to test if number of background applications affect the DNN model performance.

%\input{our_approach}
%\input{model}
%\input{algorithm}
\section{Measurement Results}\label{sec:results}

This section presents our measurement results.
The key findings are: (1) background apps
have little impact on the inference latency;
and (2) CPU contention results in a huge inference
latency variability. Interestingly,
not all DNN models are equally robust against
CPU contention than others:
 a model that has a similar latency performance
with another model given mild CPU contention
can become much worse than the other model when the CPU contention becomes more severe.
Moreover, two models exhibiting similar latencies
on one device can behave very differently on another device.

%\begin{figure}[!t]
%	\centering
%	%\subfigure[MobileNet V1Q]{\label{fig:S5_Mobilenet_V1_quant_fitting}\includegraphics[trim=0 0 0 0, clip, width=0.24\textwidth,clip]{S5_Mobilenet_V1_quant_fitting}}
%	\subfigure[MobileNet V2Q]{\label{fig:S5_Mobilenet_V2_quant_fitting}\includegraphics[trim=0 0 0 0, clip, width=0.23\textwidth,clip]{S5_Mobilenet_V2_quant_fitting}}
%%\subfigure[MobileNet V1F]{\label{fig:S5_Mobilenet_V1_float_fitting}\includegraphics[trim=0 0 0 0, clip, width=0.24\textwidth,clip]{S5_Mobilenet_V1_float_fitting}}
%\subfigure[MobileNet V2F]{\label{fig:S5_Mobilenet_V2_float_fitting}\includegraphics[trim=0 0 0 0, clip, width=0.23\textwidth,clip]{S5_Mobilenet_V2_float_fitting}}
%\squeezeup
%	\caption{Inference latency  vs. CPU usage for different MobileNet models on Samsung Galaxy	Tab	S5e}\label{fig:fitting_MobileNet_S5}
%\end{figure}

\subsection{Impact of Background Apps}

We first show in Fig.~\ref{fig:background_app} the latencies of two different DNN models on the two mobile devices, under different numbers of background apps.
It can be seen that while latency variability inevitably exists,
it is rather minimum. For example, Inception V3F exhibits a $<5\%$ variability
on the two mobile devices. Although MobileNet V2Q has a relatively larger
variability in percentage, its absolute variability is still small.
The results for the other DNN models on these devices are similar and hence omitted
for brevity.
The little inference latency variability with respect
to the number of background apps is partly attributed
to the Android's resource management, which separates background apps
and foreground apps. Moreover, when put into the background mode,
the apps only keep  minimum ongoing activities and hence result
in little resource contention.

\subsection{Impact of CPU Contention}

While background apps do not create aggressive CPU
contention, we now turn to the impact CPU contention
on latency variability  created by concurrent threads within the same 
foreground app.

We first show in Fig.~\ref{fig:thread_trace} a snapshot of CPU/memory usage and latencies
of  MobileNet V2F on Samsung S5e by gradually increasing
the number of concurrent threads (roughly every 2 minutes). We also show
the moving average latency over the most recent 20 seconds.
Since our launched concurrent
threads are all computing-intensive, the memory usage does not noticeably
vary when we increase the number of concurrent threads.
As expected,
with more concurrent threads,  the CPU usage increases
and so does the inference latency. Moreover, the latency variability
is also significant, differing from
the case in which we only increase the number of background apps
that do not utilize CPU resources.

%We also plot in Fig.~\ref{fig:fitting_MobileNet_S5} the latency vs. CPU usage for MobileNet V2Q and V2F
%on Samsung S5e. It can be seen that even with the same CPU
%usage, the inference latency can vary significantly

\begin{figure*}[!t]
	\centering
	\subfigure[0 thread]{\label{fig:S5_Mobilenet_Thread_0}\includegraphics[trim=0 0 0 0, clip, width=0.19\textwidth,clip]{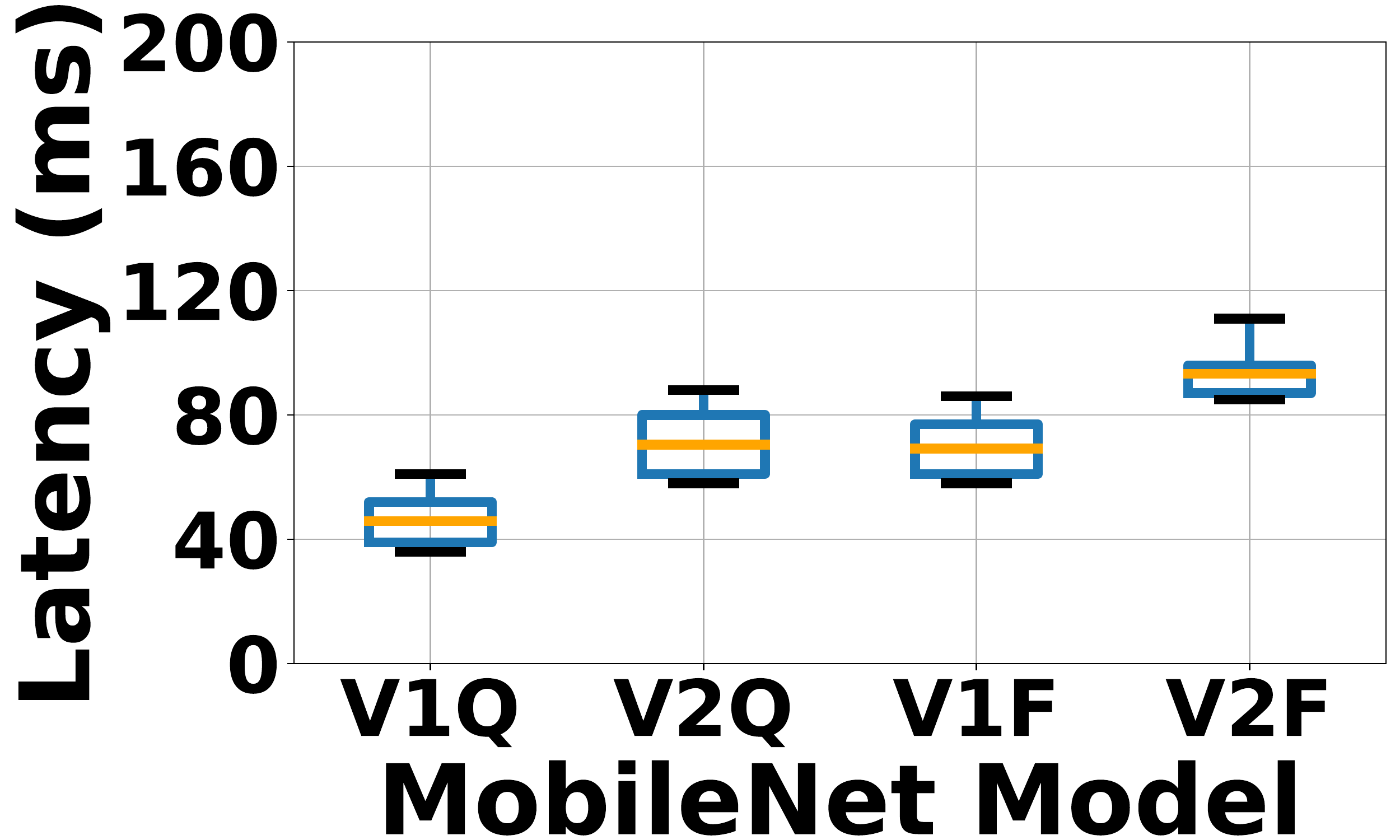}}
	\subfigure[2 threads, 20\% active]{\label{fig:S5_Mobilenet_Thread_2}\includegraphics[trim=0 0 0 0, clip, width=0.19\textwidth,clip]{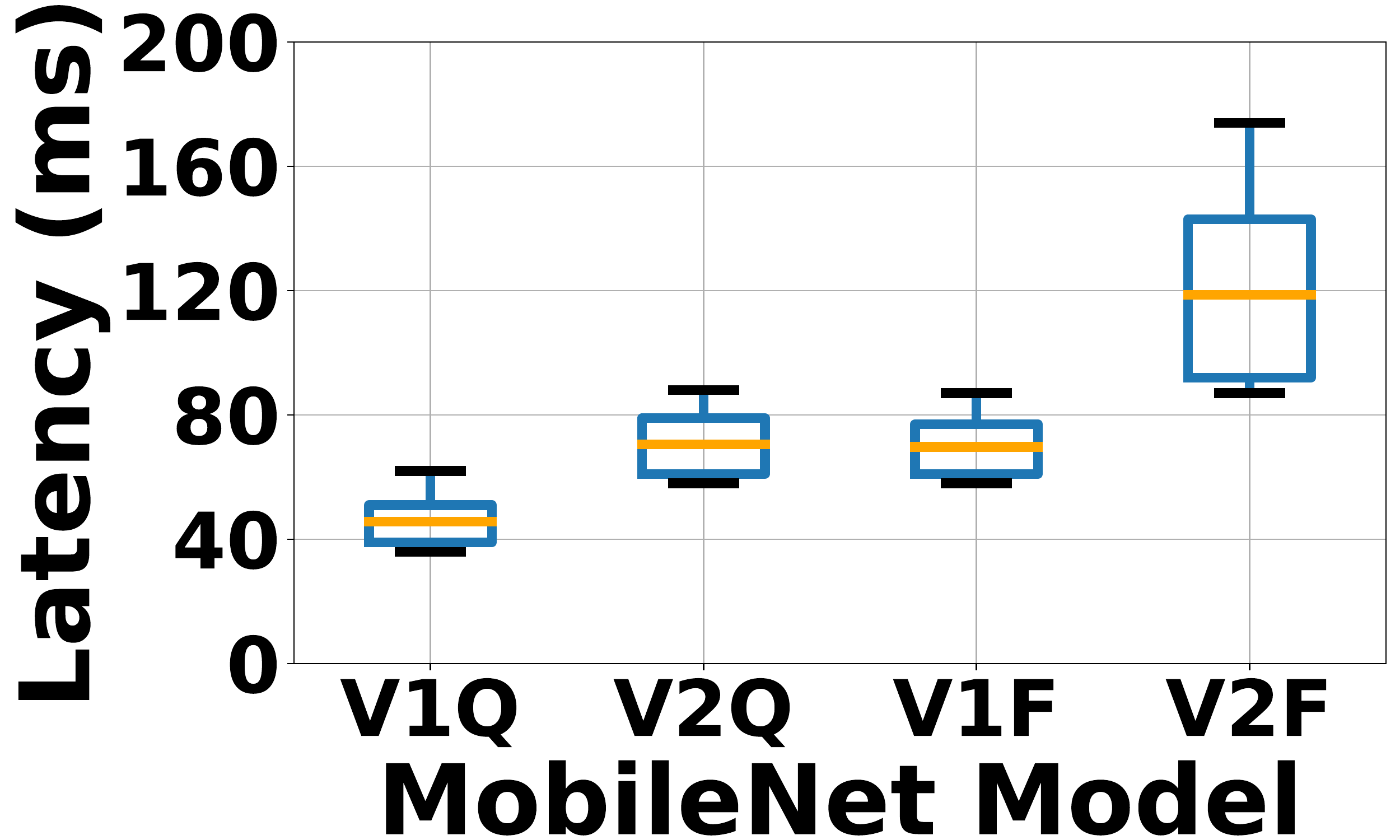}}
\subfigure[4 threads, 40\% active]{\label{fig:S5_Mobilenet_Thread_4}\includegraphics[trim=0 0 0 0, clip, width=0.19\textwidth,clip]{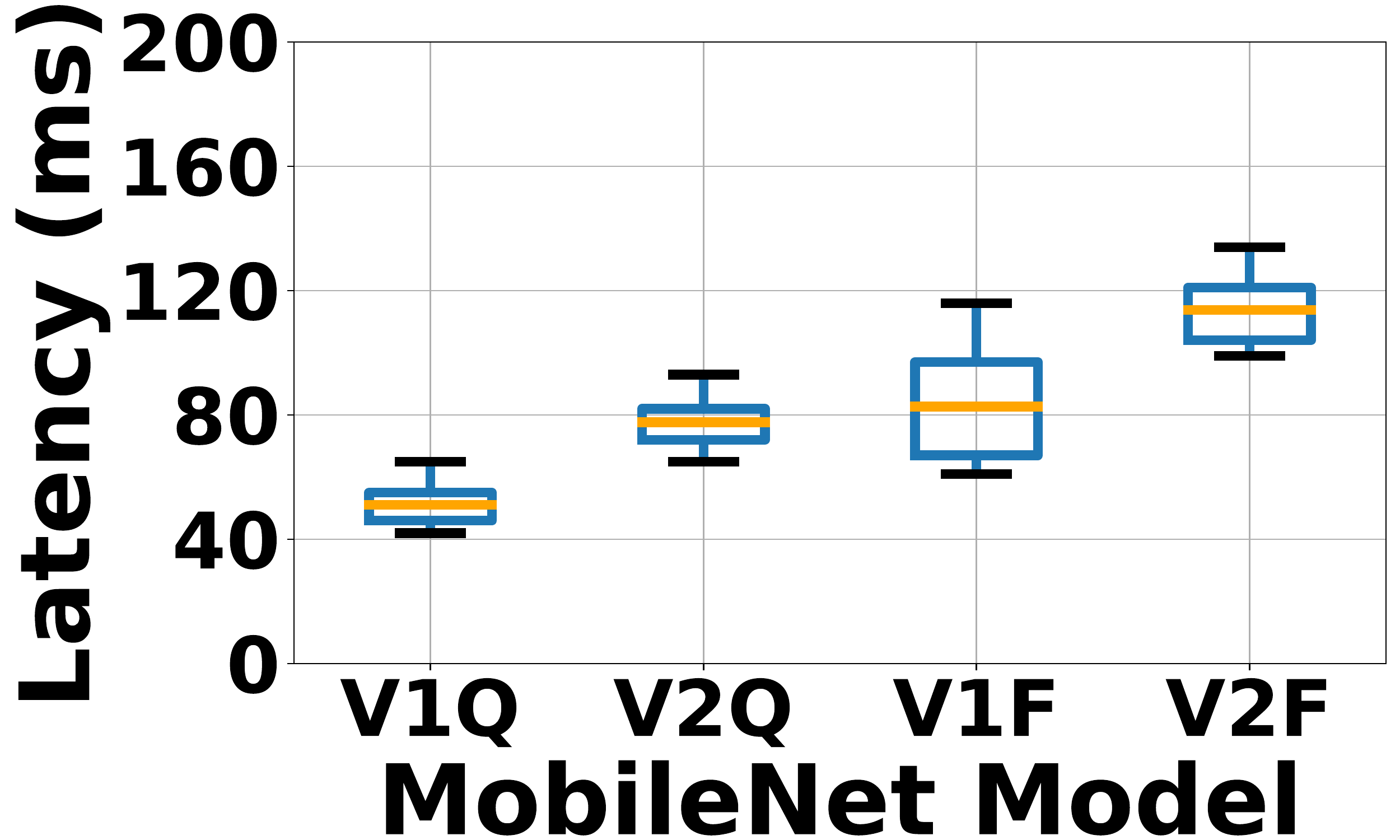}}
\subfigure[6 threads, 60\% active]{\label{fig:S5_Mobilenet_Thread_6}\includegraphics[trim=0 0 0 0, clip, width=0.19\textwidth,clip]{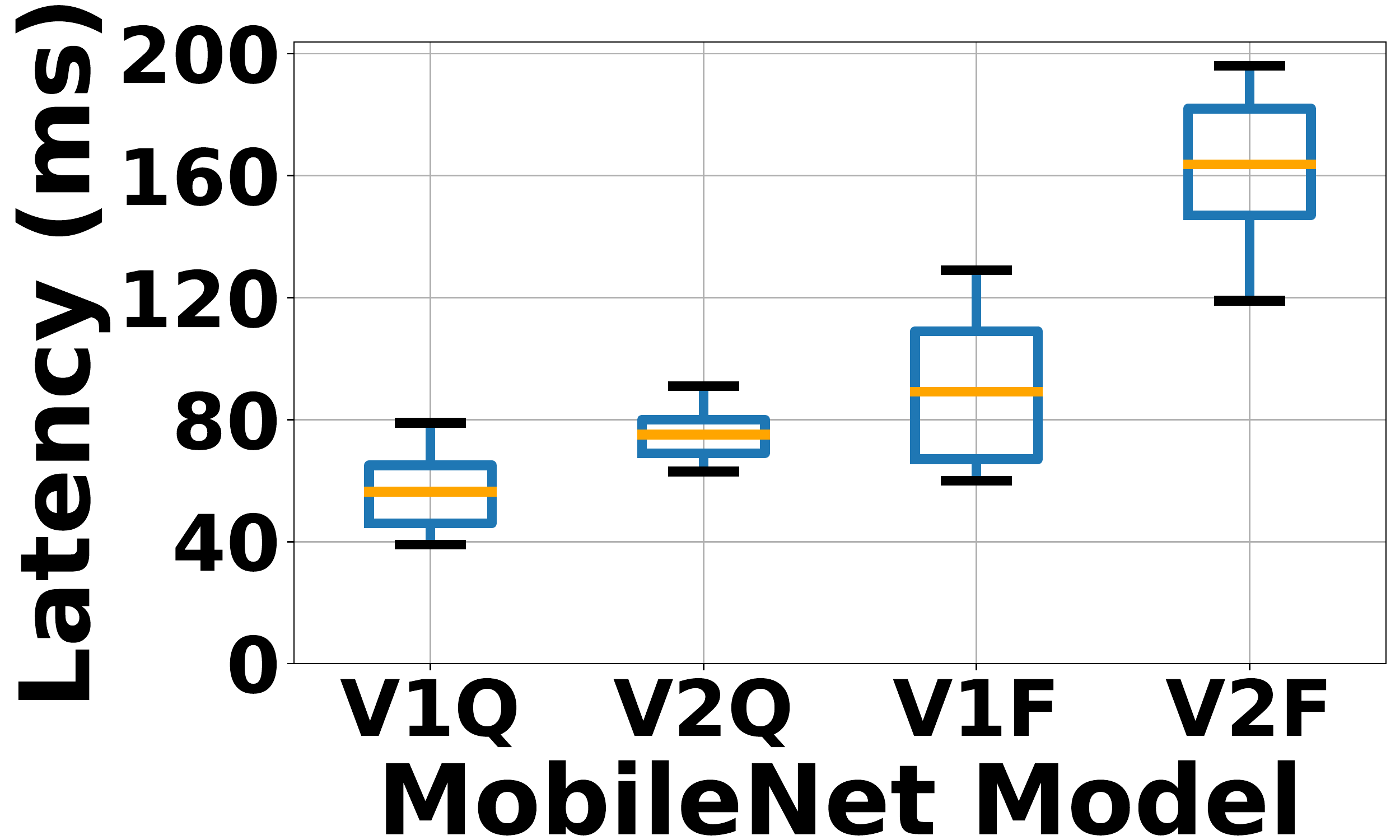}}
\subfigure[8 threads, 80\% active]{\label{fig:S5_Mobilenet_Thread_8}\includegraphics[trim=0 0 0 0, clip, width=0.19\textwidth,clip]{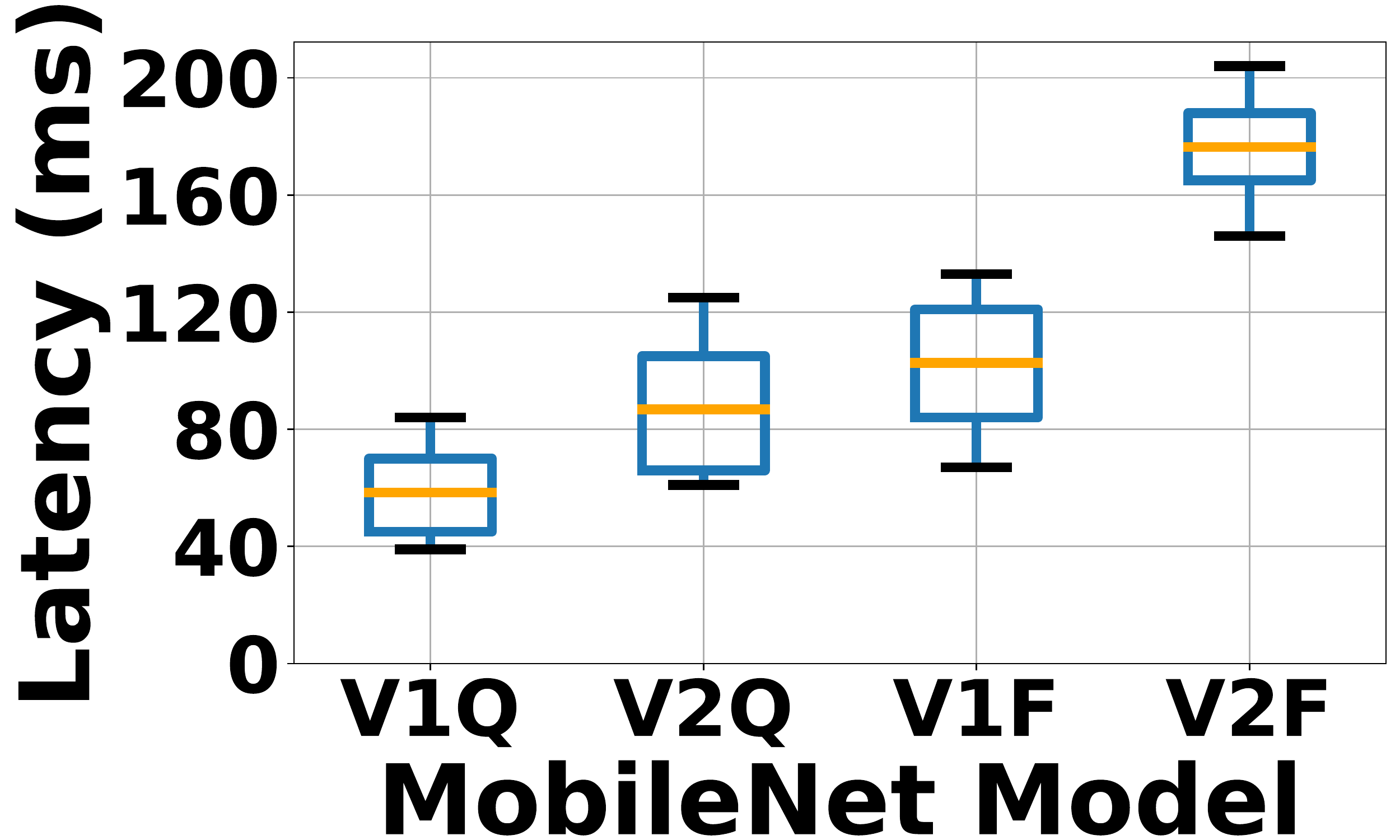}}
\squeezeup
	\caption{Inference latency of MobileNet  on Samsung	Galaxy	Tab	S5e with different numbers of concurrent threads.}\label{fig:thread_S5_Mobilenet}
\end{figure*}

\begin{figure*}[!t]
	\centering
	\subfigure[0 thread]{\label{fig:A_Mobilenet_Thread_0}\includegraphics[trim=0 0 0 0, clip, width=0.19\textwidth,clip]{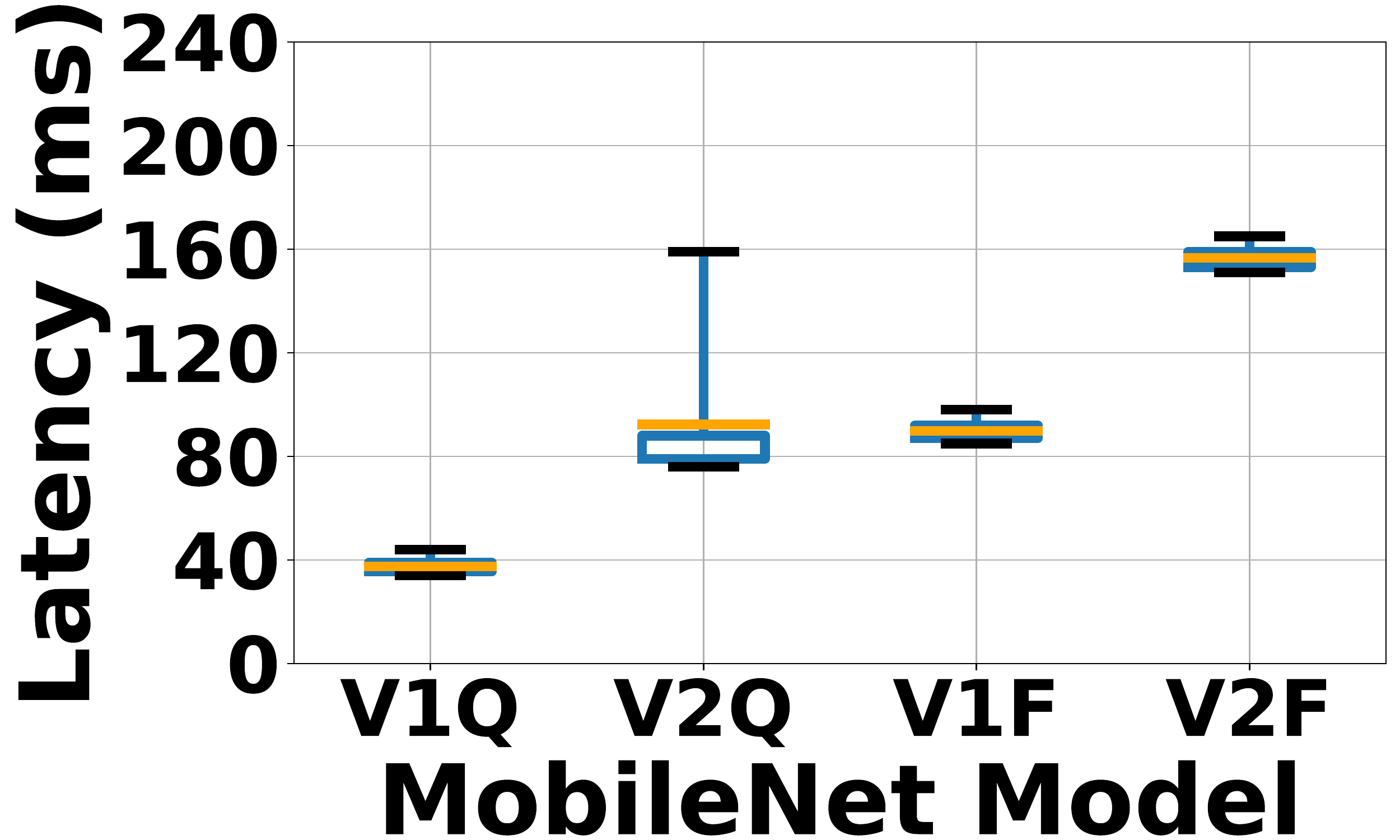}}
	\subfigure[2 threads, 20\% active]{\label{fig:A_Mobilenet_Thread_2}\includegraphics[trim=0 0 0 0, clip, width=0.19\textwidth,clip]{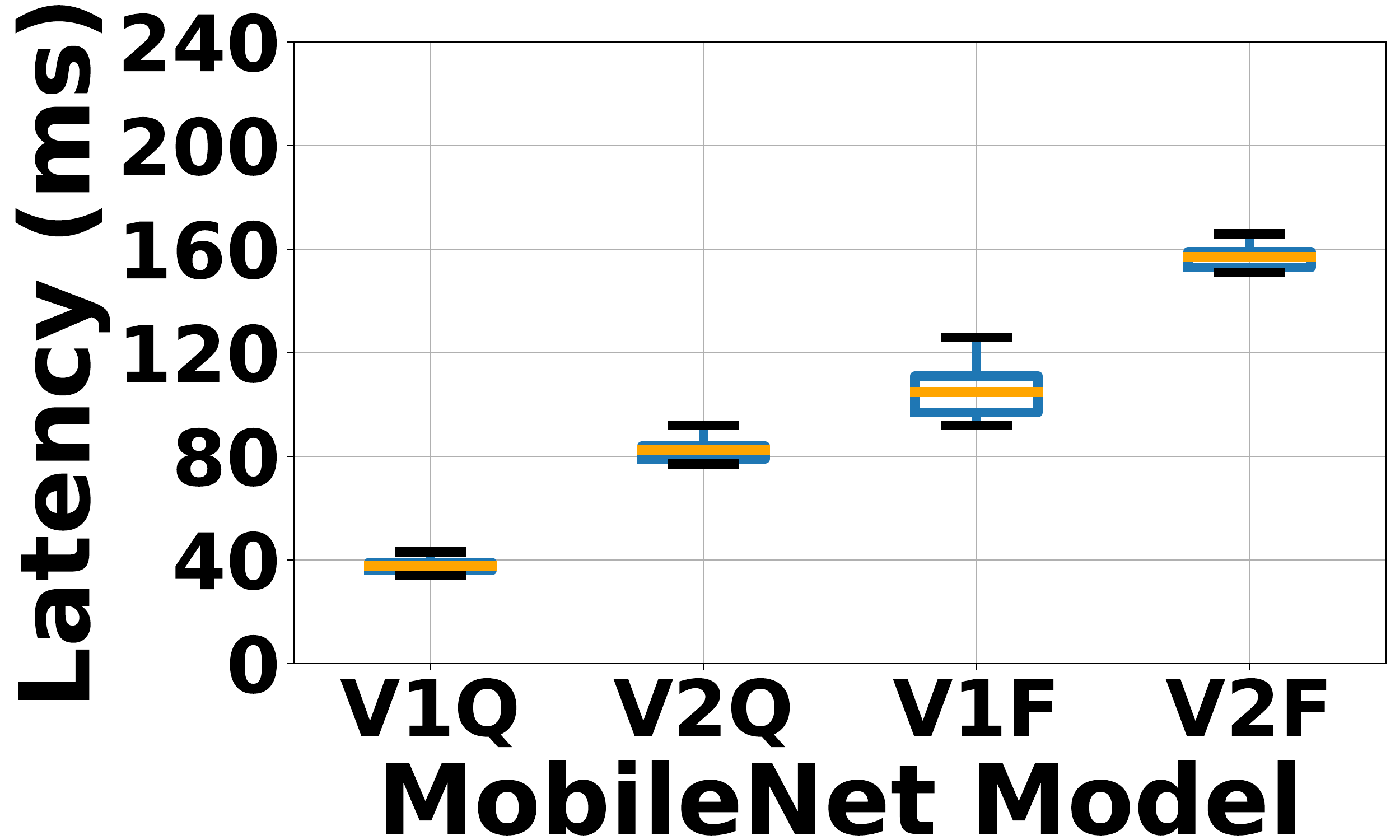}}
\subfigure[4 threads, 40\% active]{\label{fig:A_Mobilenet_Thread_4}\includegraphics[trim=0 0 0 0, clip, width=0.19\textwidth,clip]{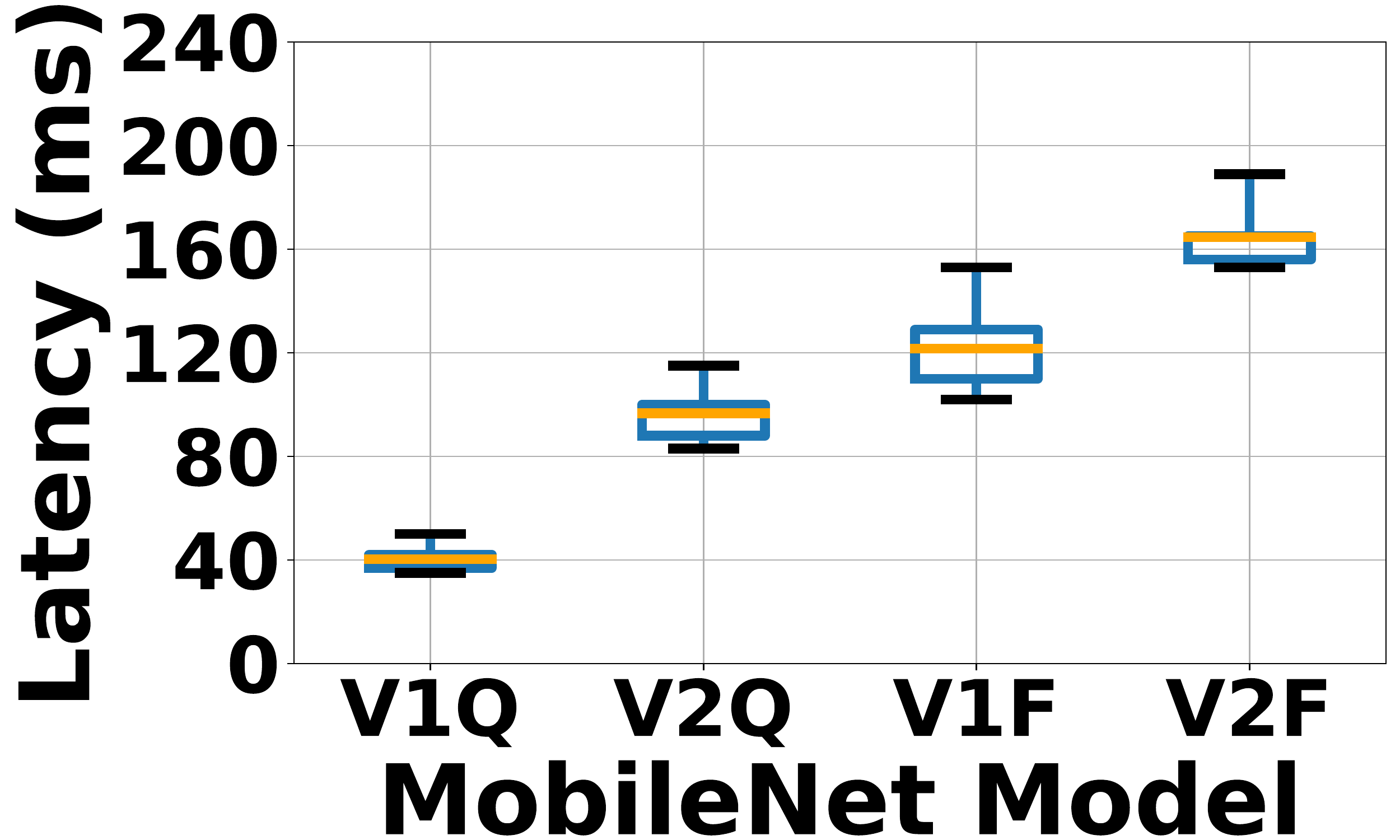}}
\subfigure[6 threads, 60\% active]{\label{fig:A_Mobilenet_Thread_6}\includegraphics[trim=0 0 0 0, clip, width=0.19\textwidth,clip]{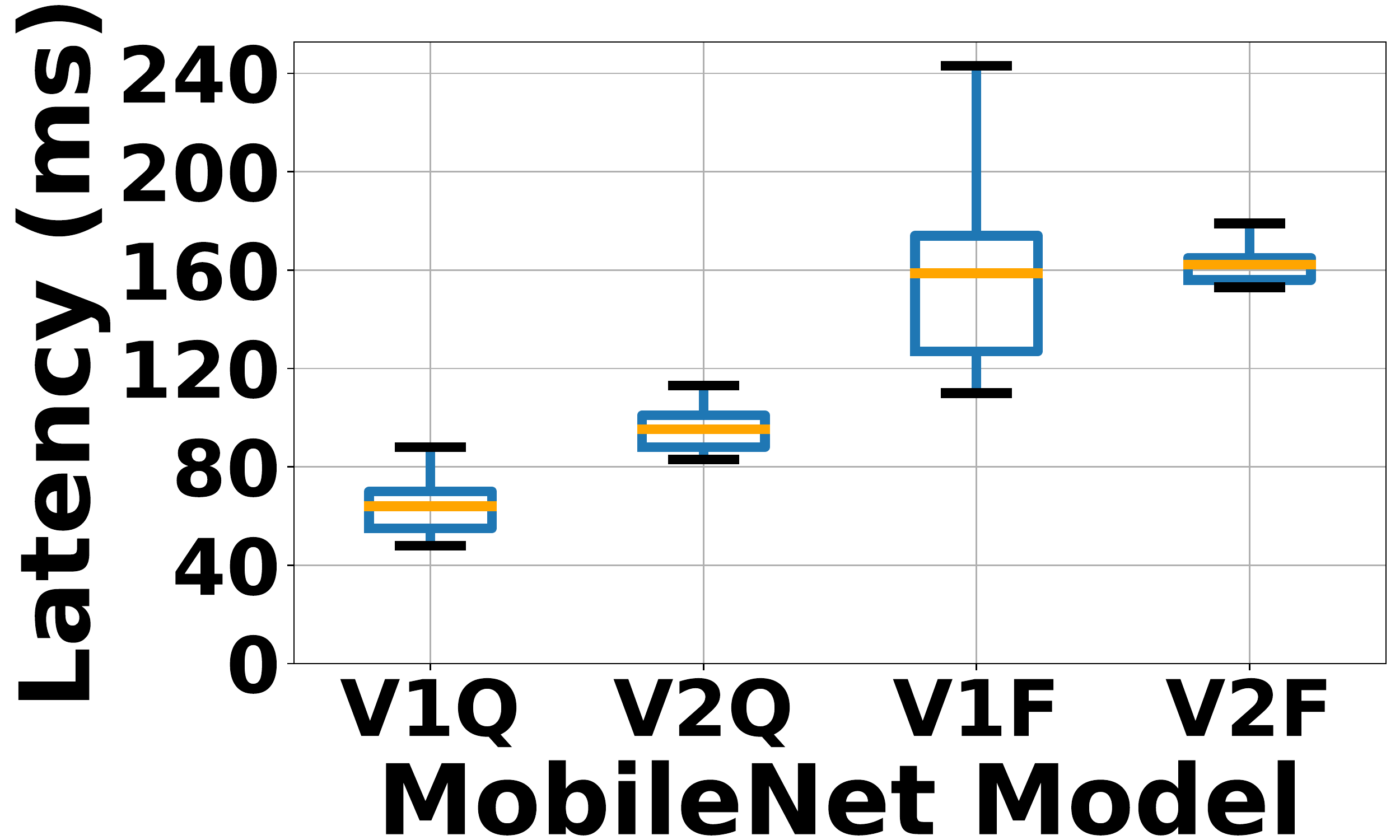}}
\subfigure[8 threads, 80\% active]{\label{fig:A_Mobilenet_Thread_8}\includegraphics[trim=0 0 0 0, clip, width=0.19\textwidth,clip]{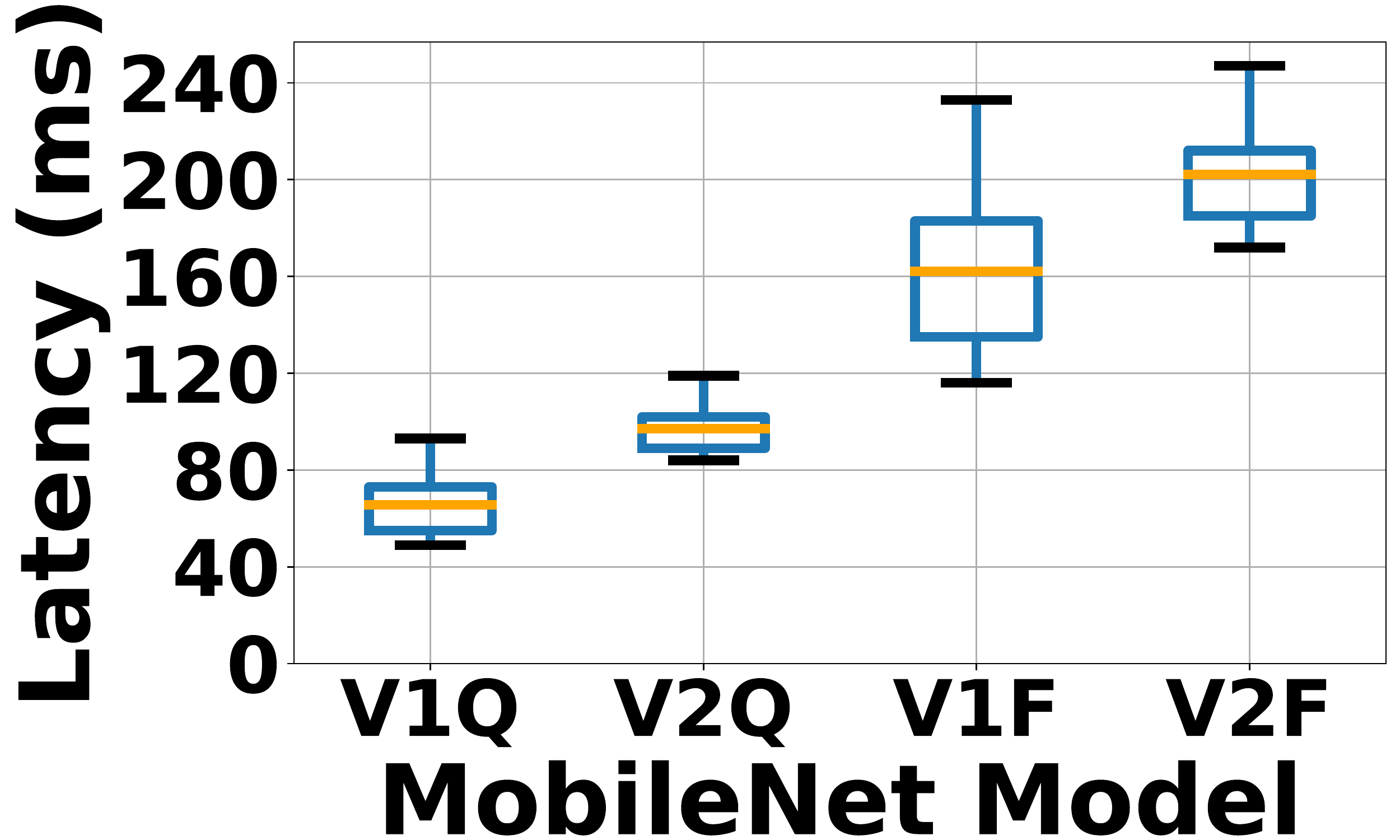}}
\squeezeup
	\caption{Inference latency of MobileNet  on Samsung	Galaxy	Tab	A with different numbers of concurrent threads.}\label{fig:thread_A_Mobilenet}
\end{figure*}

\begin{figure*}[!t]
	\centering
	\subfigure[0 thread]{\label{fig:S5_Inception_Thread_0}\includegraphics[trim=0 0 0 0, clip, width=0.19\textwidth,clip]{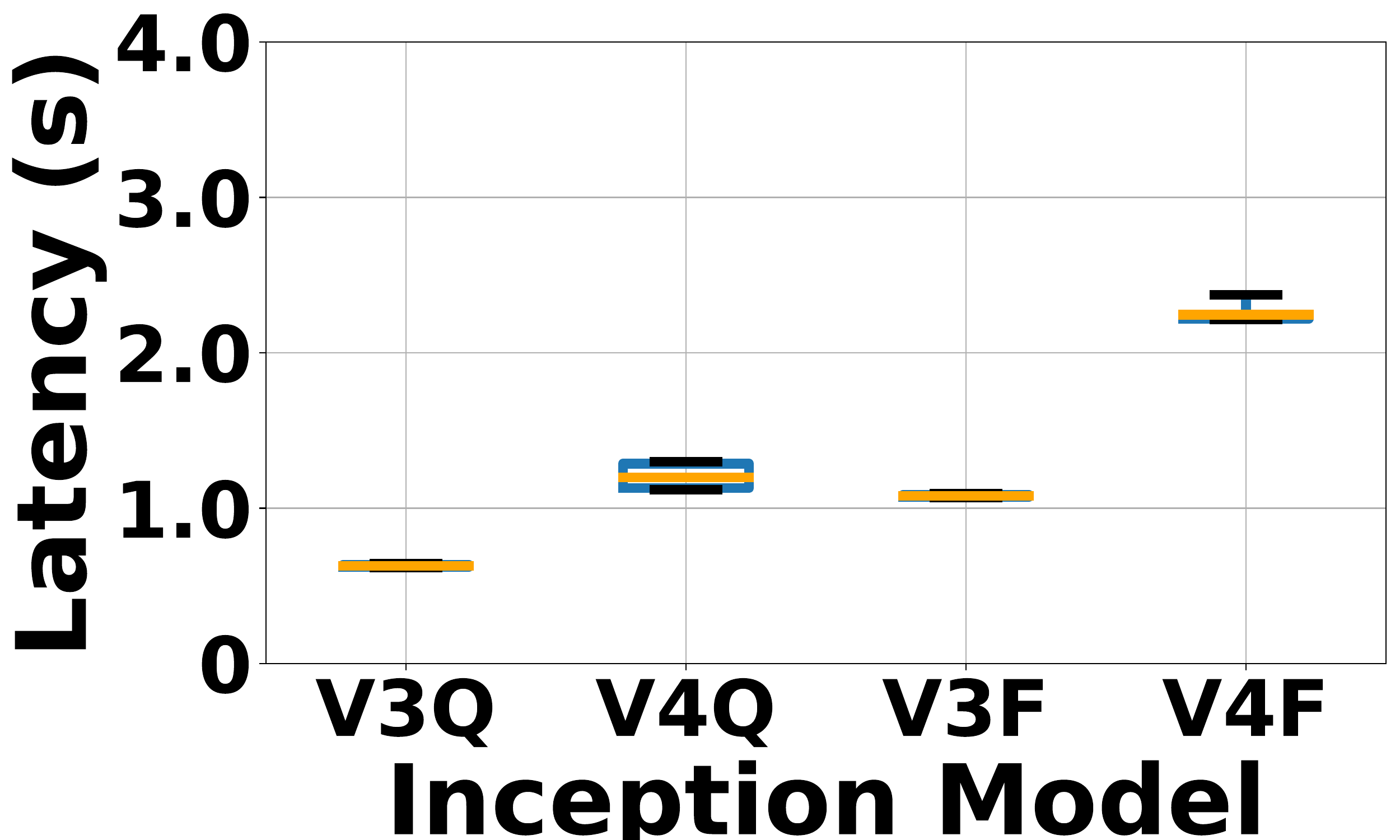}}
	\subfigure[2 threads, 20\% active]{\label{fig:S5_Inception_Thread_2}\includegraphics[trim=0 0 0 0, clip, width=0.19\textwidth,clip]{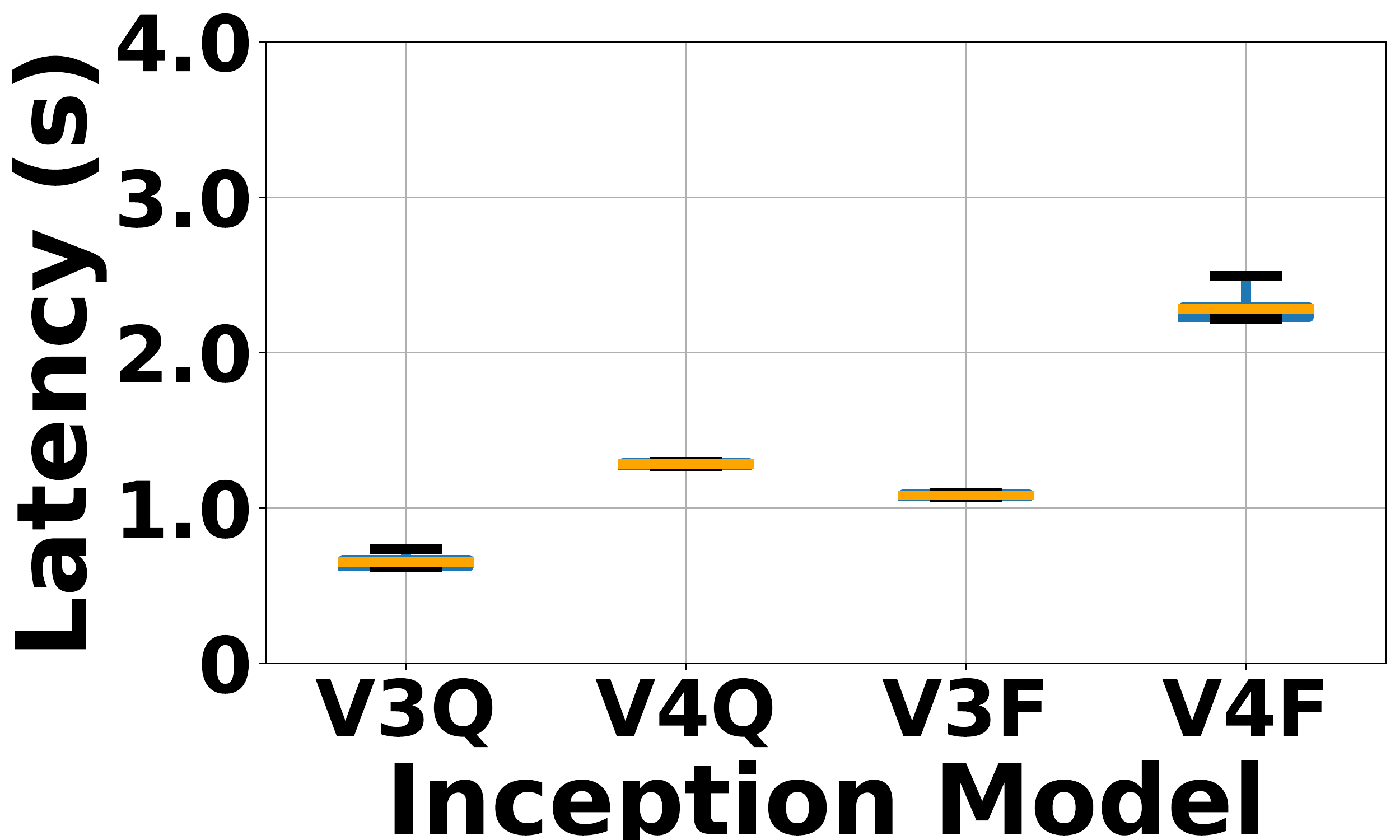}}
\subfigure[4 threads, 40\% active]{\label{fig:S5_Inception_Thread_4}\includegraphics[trim=0 0 0 0, clip, width=0.19\textwidth,clip]{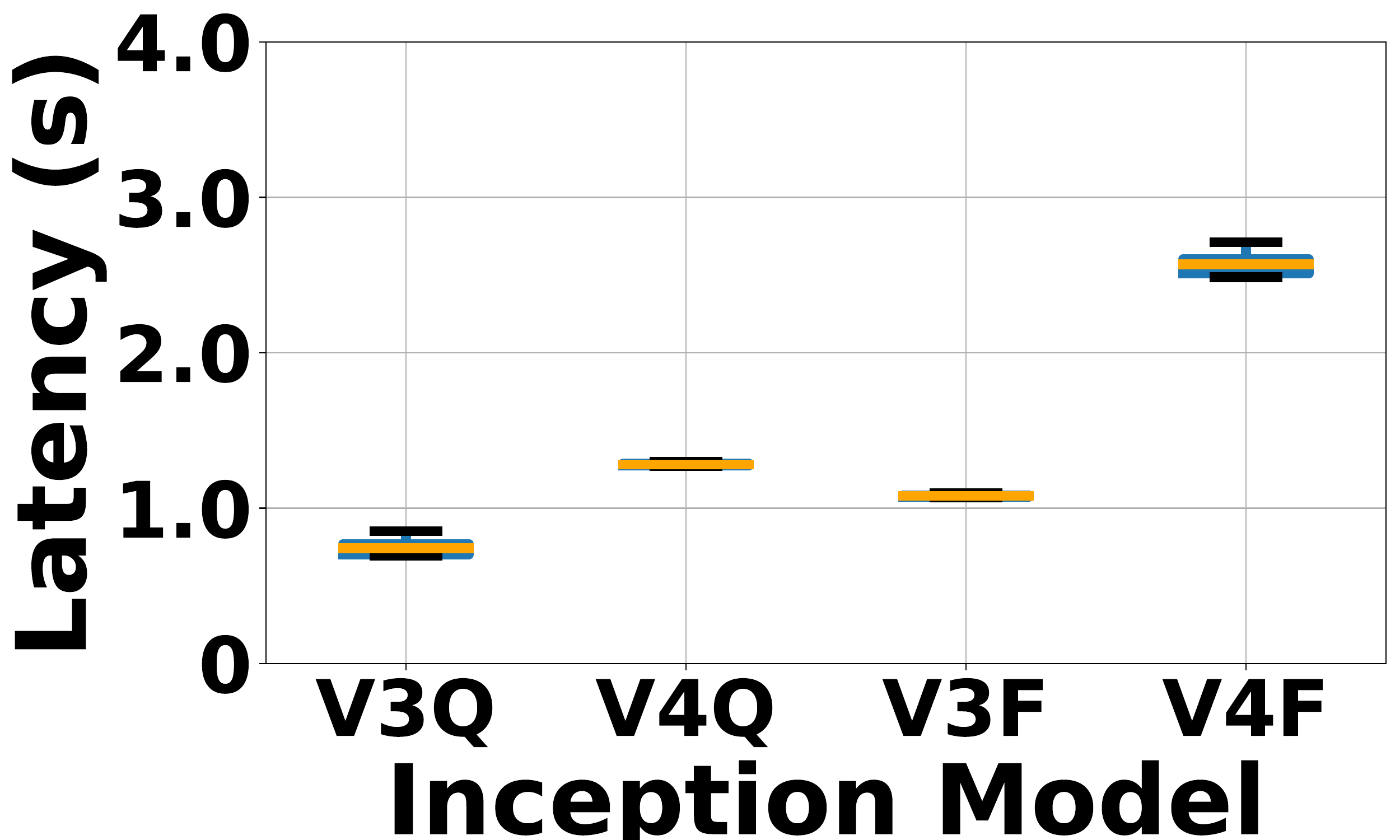}}
\subfigure[6 threads, 60\% active]{\label{fig:S5_Inception_Thread_6}\includegraphics[trim=0 0 0 0, clip, width=0.19\textwidth,clip]{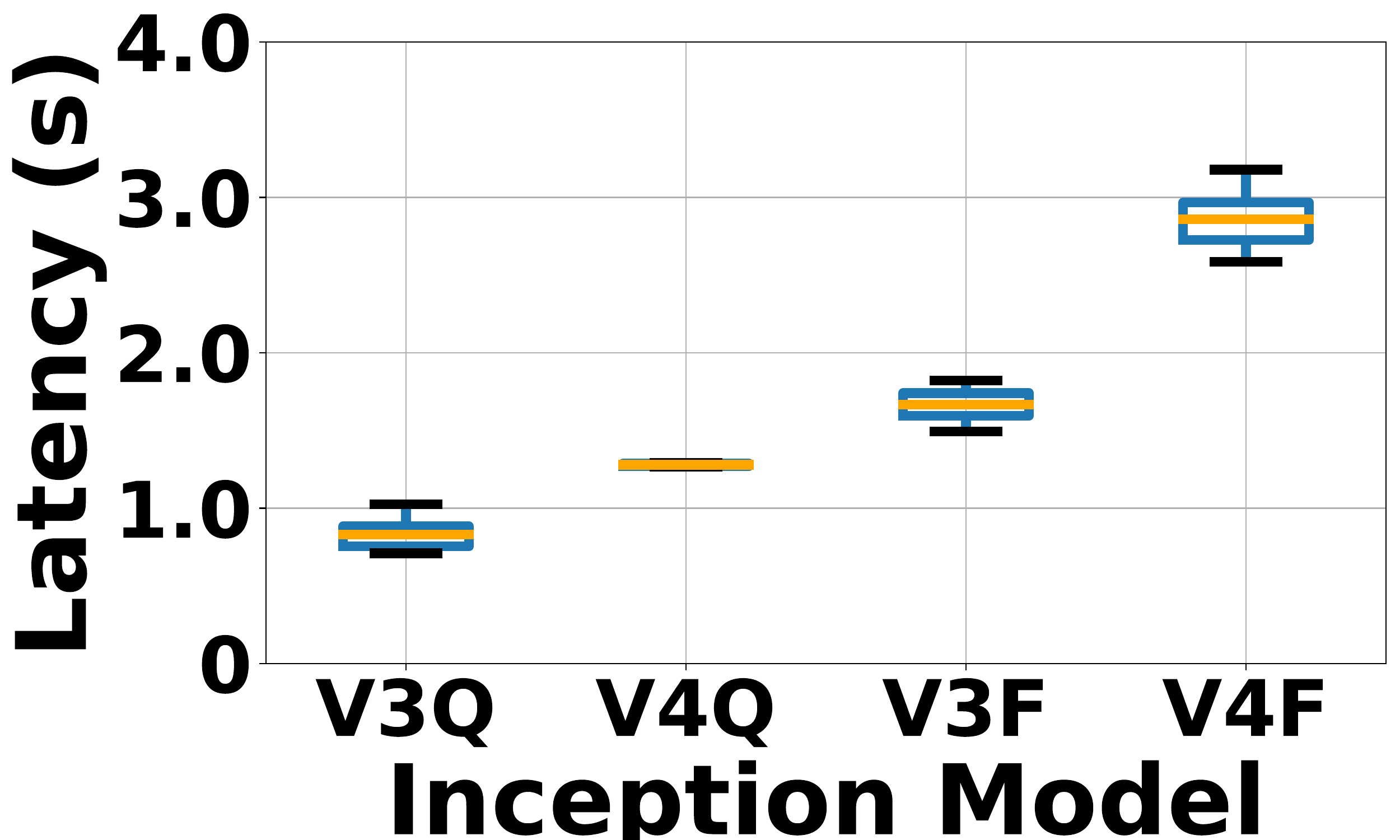}}
\subfigure[8 threads, 80\% active]{\label{fig:S5_Inception_Thread_8}\includegraphics[trim=0 0 0 0, clip, width=0.19\textwidth,clip]{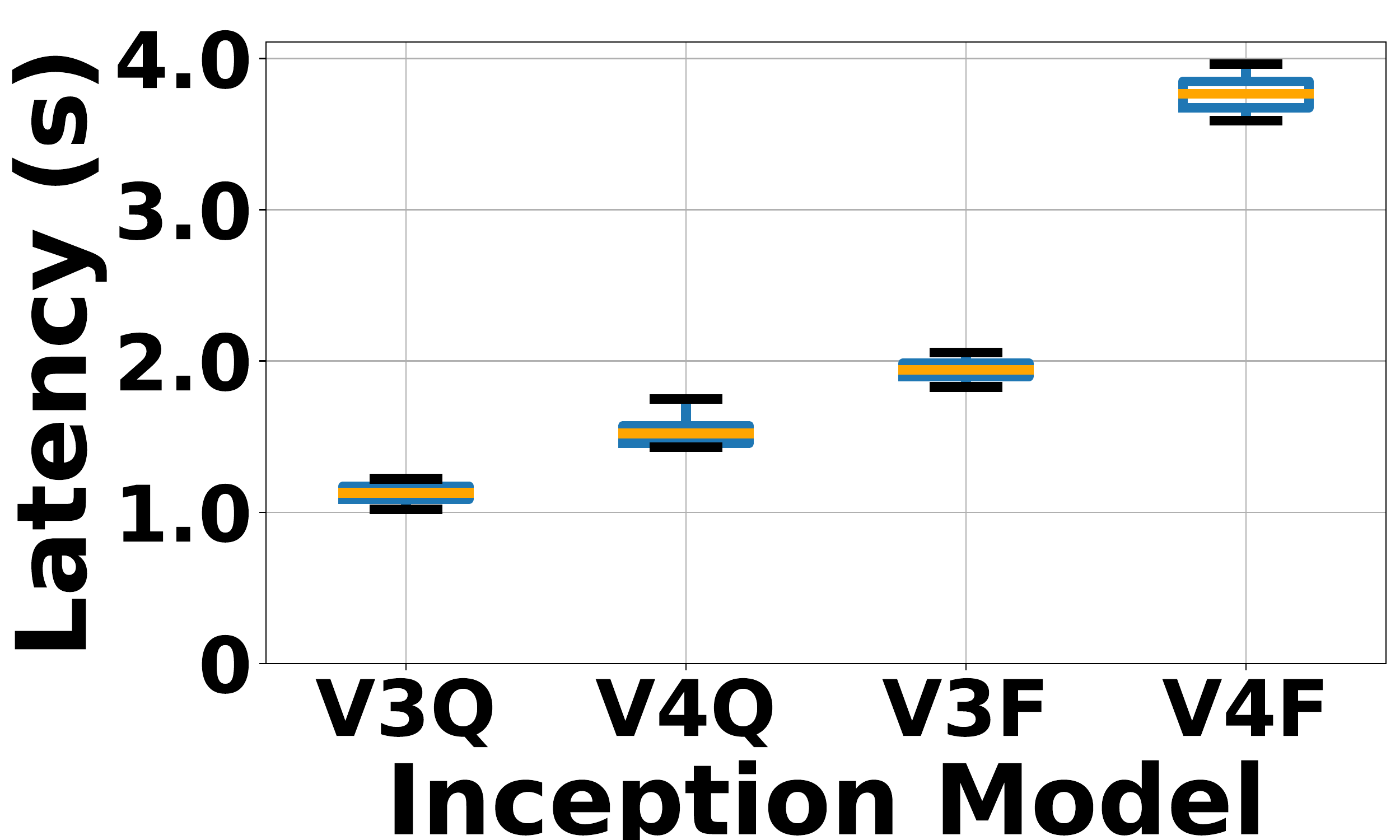}}
\squeezeup
	\caption{Inference latency of Inception  on Samsung	Galaxy	Tab	S5e with different numbers of concurrent threads.}\label{fig:thread_S5_Inception}
\end{figure*}

\begin{figure*}[!t]
	\centering
	\subfigure[0 thread]{\label{fig:A_Inception_Thread_0}\includegraphics[trim=0 0 0 0, clip, width=0.19\textwidth,clip]{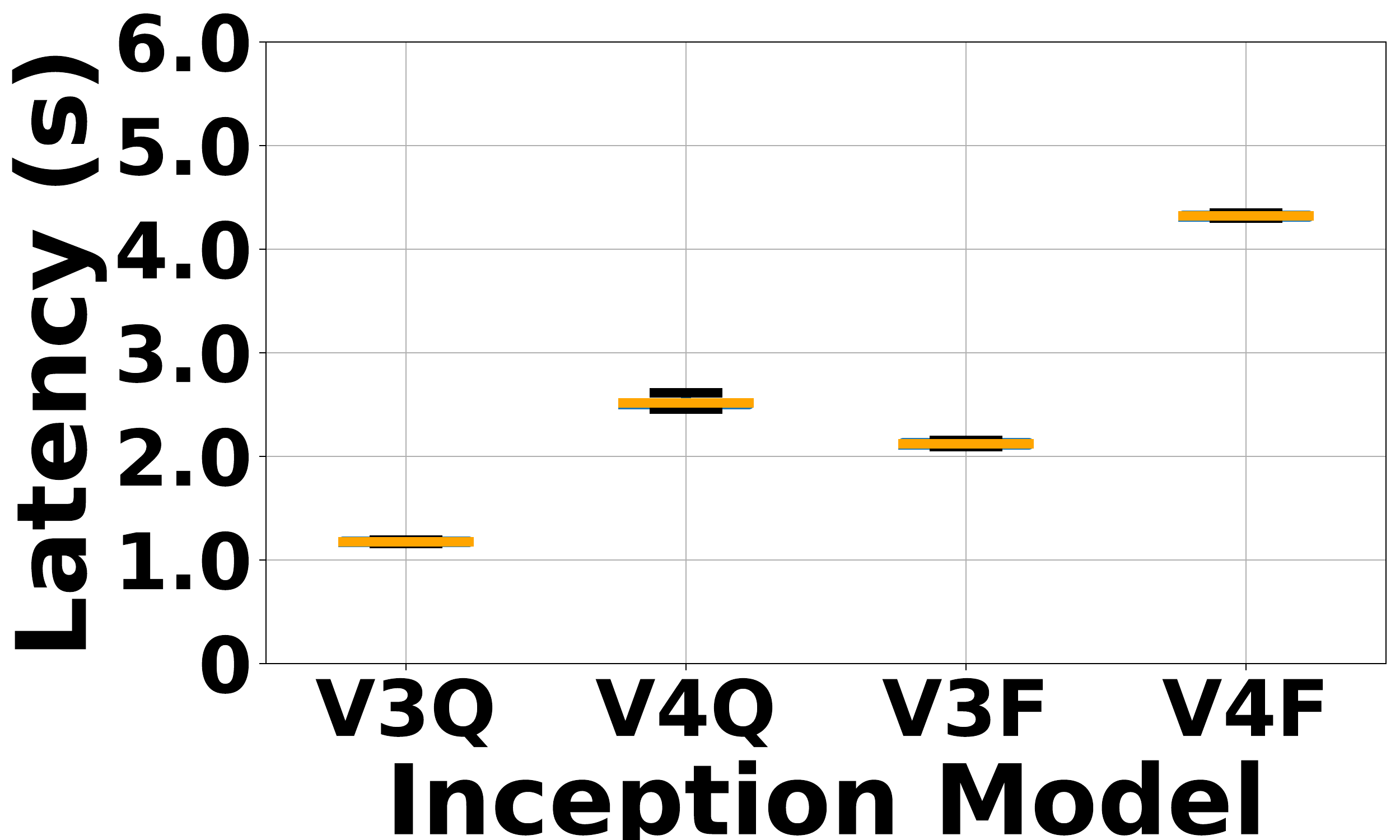}}
	\subfigure[2 threads, 20\% active]{\label{fig:A_Inception_Thread_2}\includegraphics[trim=0 0 0 0, clip, width=0.19\textwidth,clip]{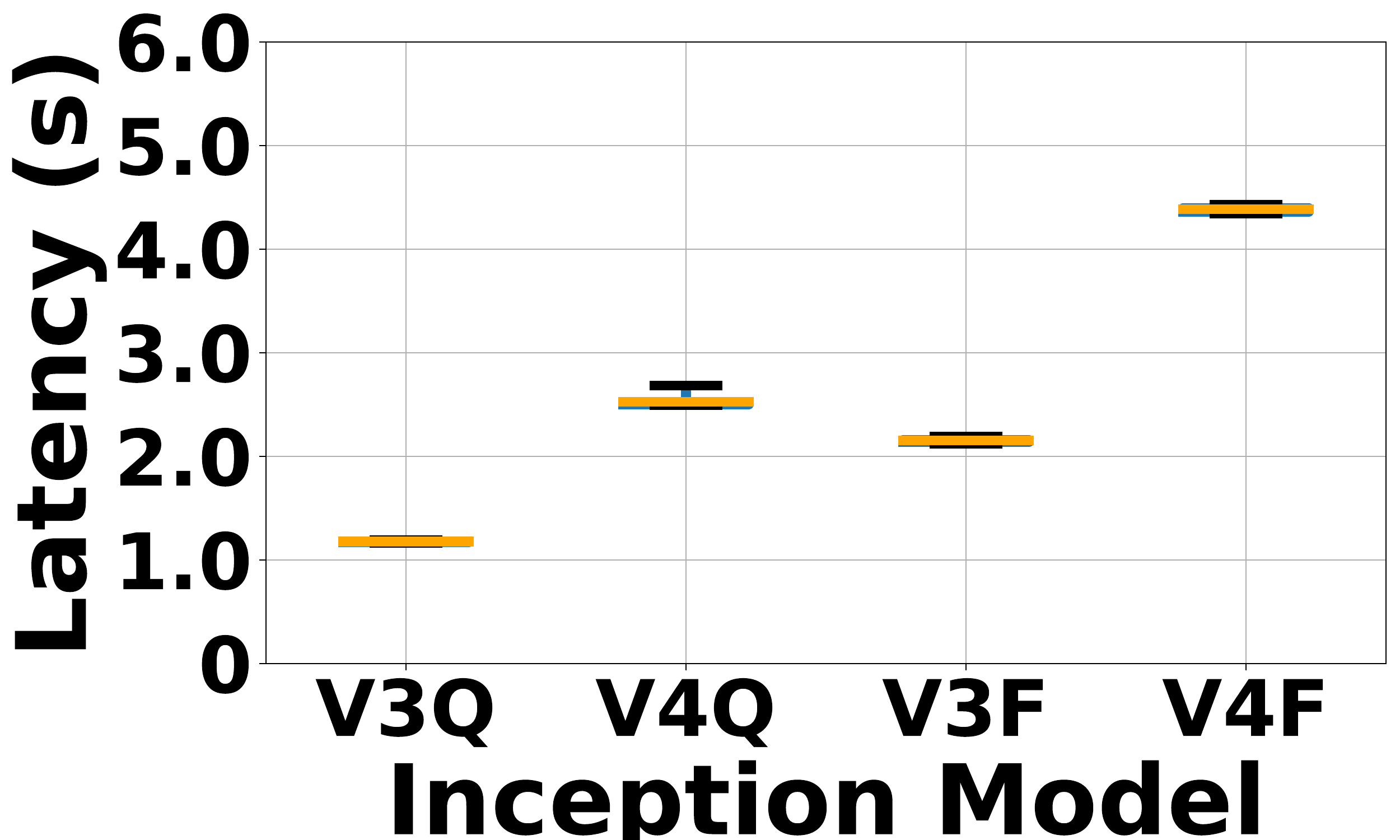}}
\subfigure[4 threads, 40\% active]{\label{fig:A_Inception_Thread_4}\includegraphics[trim=0 0 0 0, clip, width=0.19\textwidth,clip]{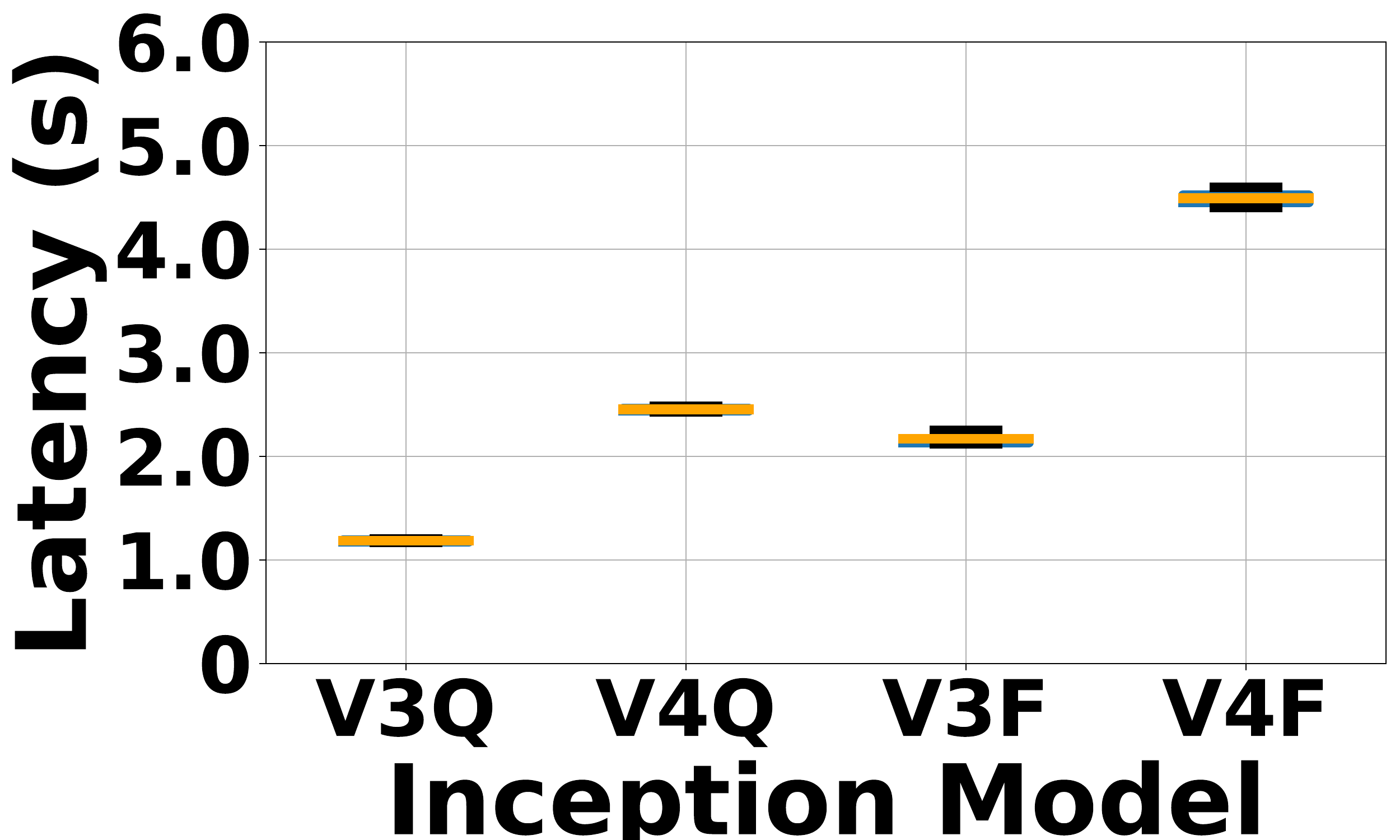}}
\subfigure[6 threads, 60\% active]{\label{fig:A_Inception_Thread_6}\includegraphics[trim=0 0 0 0, clip, width=0.19\textwidth,clip]{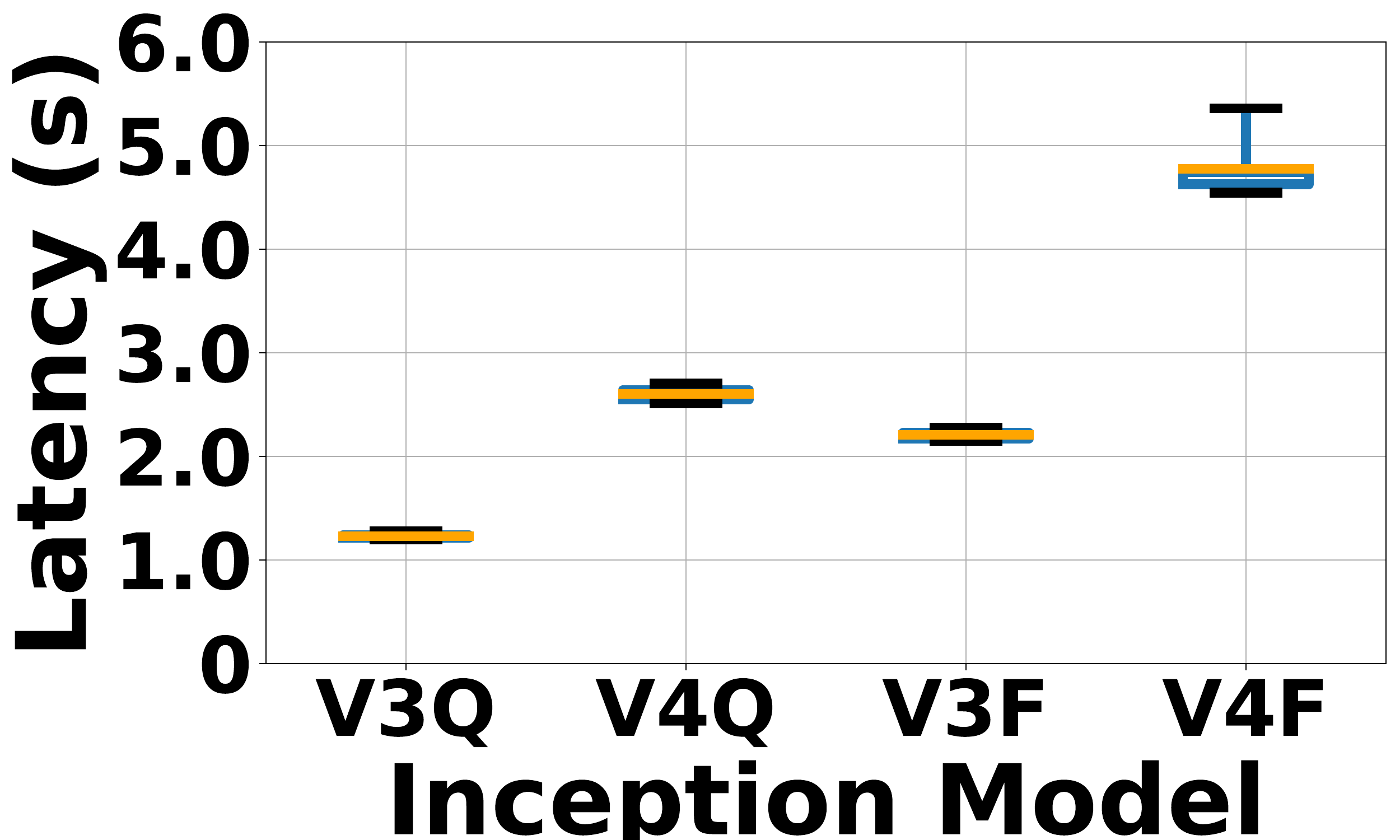}}
\subfigure[8 threads, 80\% active]{\label{fig:A_Inception_Thread_8}\includegraphics[trim=0 0 0 0, clip, width=0.19\textwidth,clip]{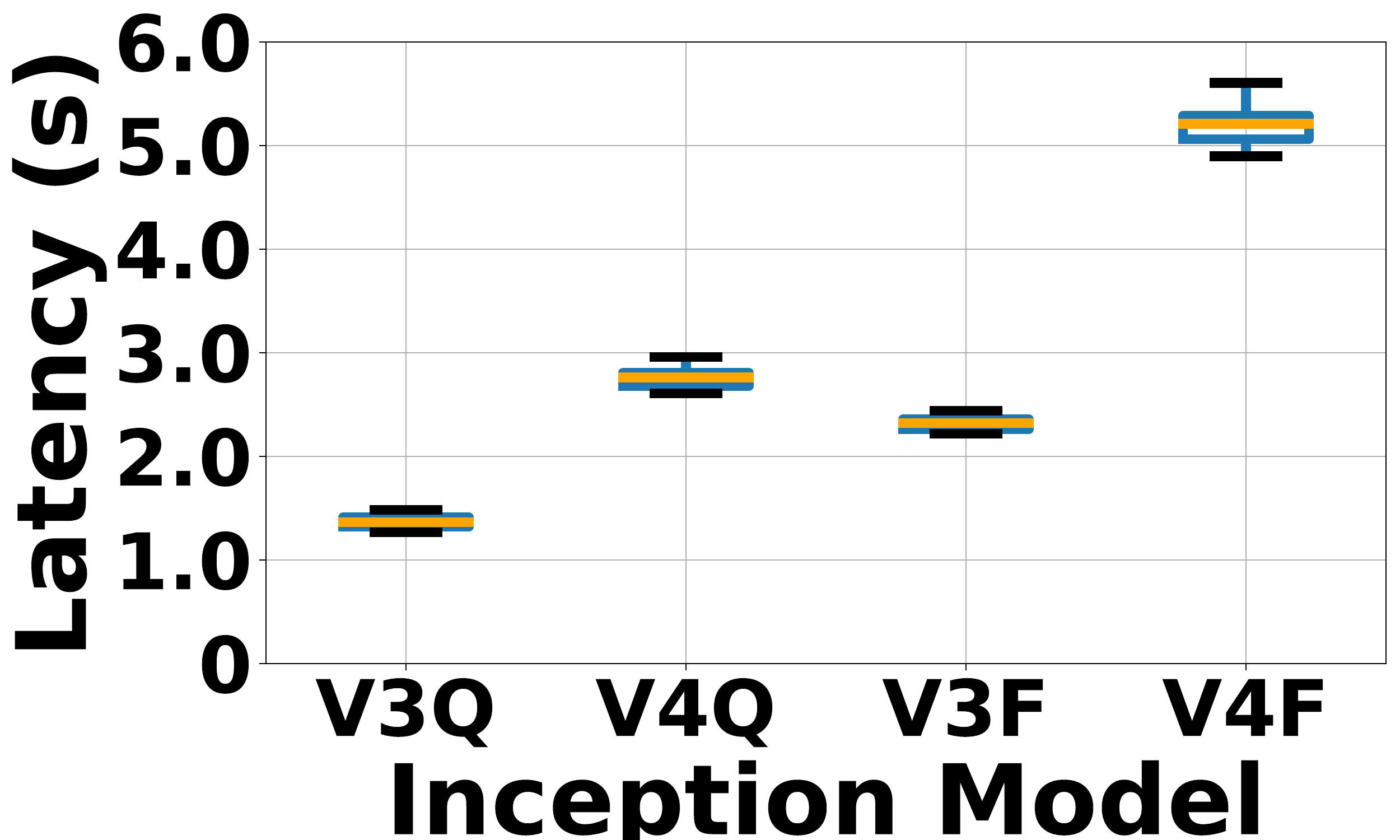}}
\squeezeup
	\caption{Inference latency of Inception  on Samsung	Galaxy	Tab	A with different numbers of concurrent threads.}\label{fig:thread_A_Inception}
\end{figure*}

Next, from Fig.~\ref{fig:thread_S5_Mobilenet}
to Fig.~\ref{fig:thread_A_Inception}, we show the latency measurement
results for eight DNN models on two mobile devices, under
different numbers of concurrent threads.  Note that as in
other error-bar plots, we show the 5th, 25th, 75th and 95th percentile
and average latencies, excluding top and bottom 5\% latencies.
The ``$x\%$ active'' in the figure captions
indicates each thread has a probability of $x\%$ to perform computation
for each time slot of 5ms. We can observe the following.

$\bullet$ First, for any given DNN model running on a device,
the average latency increases with more CPU contention created
by more concurrent threads. %This is intuitively expected.
Nonetheless, not all DNN models have the same amount of latency
variability increase. For example, the latency of MobileNet V1Q
increases but still exhibits a fairly small variability on
both devices. On the other hand, the latency variability
of MobileNet V1F becomes significantly larger with more CPU contention.
This shows that given the same device, different DNN models have
different robustness in terms of latency variability.

$\bullet$  Second, given the same  device, one DNN model that has a similar latency performance
with another model in the event of low CPU contention
can become much worse than the other model when the CPU contention becomes higher.
For example, when running on Samsung Tab A,
MobileNet V1F has a similar or better latency performance
compared to V2Q when the number of concurrent threads is less than 4,
but the latency performance of V1F becomes significantly worse
than V2Q when more concurrent threads are launched.
We focus on the comparison between V1F and V2Q, because they have
similar inference accuracies shown in Table~\ref{table:model_details}. While MoibleNet V1Q and V2F
have significantly different latency performances whose relative
superiority remains unchanged under different conditions,
these two models have very different accuracies. Thus, using
V1Q vs. V2F depends on how one weighs the inference accuracy and latency.
The same observation can also be made for
Inception V3F and V4Q on Samsung Tab S5e.
This implies that %when optimizing DNNs for mobile inference, 
an improved performance
of average latency and latency variability under one condition does not mean
the model will always have better performance under another condition.
Thus, our results highlight the need of considering different runtime conditions
when optimizing DNN models for mobile inference.

$\bullet$  Last but not least, two models exhibiting similar latencies
on one device can behave very differently on another device.
The existing research on DNN model optimization for mobile inference
typically considers a small number of or only one mobile device
and reports the average latency, implicitly assuming
that the better latency performance of DNN model on one device
will translate into a better performance on another device
(although the absolute latency will vary depending on the actual device).
Nonetheless, our measurement results invalidate this assumption.
For example, when running on Samsung Tab S5e with  two concurrent threads, MobileNet V1F
is comparable to or better than V2Q in terms of latency, but
V2Q becomes better than V1F in terms of latency and variability when running on Samsung Tab A. The same observation
can be made for Inception V3F and V4Q when running on Samsung Tab S5e and Tab A, respectively, under six or eight concurrent threads.
Our results highlight that the relative superiority of latency
performance of DNN models depends on the mobile device that runs the model
(due to operating system, hardware configuration, etc.).
Thus, when optimizing DNN models for mobile inference,
one should cover as many mobile devices as possible.

\section{Related Work}\label{sec:related_work}

To enable DNN deployment on resource-constraint mobile devices, various model compression methods have been proposed, including network and weight pruning \cite{han2015deep,han2015learning,lecun1990optimal,han2015learning,li2016pruning}, weight quantization \cite{courbariaux2016binarized,rastegari2016xnor}, low-rank matrix approximation \cite{denton2014exploiting,lebedev2014speeding},
 knowledge distillation \cite{hinton2015distilling},
and/or a combination of basic compression techniques \cite{srinivas2015data,molchanov2017variational,louizos2017learning}.
These studies focus on optimizing and measuring the average inference latency
in an static (and often idealized) environment with little resource contention.

Another recent study \cite{DNN_Compression_CharacterizingDNN_Mobile_Inference_TianGuo_arXiv_2019}
considers dynamically deciding between on-device mobile inference and offloading-based
inference for DNNs. While it confirms that there exists some latency variability
for mobile inference,
 it does not investigate the impact of runtime condition
(e.g., the number of concurrent CPU-intensive threads) on inference latency.
Our study explicitly focuses on the latency variability
of DNNs under different CPU contentions, which are common in practice \cite{DNN_Facebook_Inference_HPCA_2019}.

%Our work contributes both to the growing literature on data center demand response, and to the literature studying supply function equilibria. We discuss each in turn below.

%Power management in data centers has received a surging interest in the past years.
%For example, prior studies for minimizing energy cost include dynamically switching servers on/off in accordance with workloads  \cite{LinWiermanAndrewThereska,Gandhi:2012:ADR:2382553.2382556},  CPU
%frequency scaling \cite{Wierman:2012:PSS:2388127.2388334,Christos_RAPL_Lo:2014:TEP:2665671.2665718},
%geographic load balancing \cite{Liu:2011:GGL:1993744.1993767,LinLiuWiermanAndrew_IGCC_2012,RaoLiuXieLiu_2010},
%among others.

%
%To tame the high \capex for power
%and cooling infrastructure,

%To gracefully handle power emergencies,

\section{Conclusion}

In this note, we present a preliminary study
on the latency variability of DNNs for mobile inference.
While the number
of background apps has marginal impact,
the inference latency can dramatically increase with a significant
variability given more CPU contention.
 More interestingly,
 one DNN model
 with a better latency performance than another model can be
outperformed by the other model when running on another device %the level of
and/or CPU contention
becomes more severe.

Our measurement study also opens up an interesting set of questions.
Why are some DNN models are more robust against resource contention than others?
Why do the relative performance of DNN models change under
different levels of CPU contention and/or on different devices?
How to mitigate inference latency variability of DNN models for mobile inference?
{
\small
\bibliographystyle{ieeetr}
\bibliography{ref,ref_ren,ref_jie_xu,ref_bingqian}

\begin{thebibliography}{10}

\bibitem{DNN_Book_Goodfellow-et-al-2016}
I.~Goodfellow, Y.~Bengio, and A.~Courville, {\em Deep Learning}.
\newblock MIT Press, 2016.
\newblock \url{http://www.deeplearningbook.org}.

\bibitem{DNN_Facebook_Inference_HPCA_2019}
C.-J. Wu, D.~Brooks, K.~Chen, D.~Chen, S.~Choudhury, M.~Dukhan, K.~Hazelwood,
  E.~Isaac, Y.~Jia, B.~Jia, T.~Leyvand, H.~Lu, Y.~Lu, L.~Qiao, B.~Reagen,
  J.~Spisak, F.~Sun, A.~Tulloch, P.~Vajda, X.~Wang, Y.~Wang, B.~Wasti, Y.~Wu,
  R.~Xian, S.~Yoo, and P.~Zhang, ``Machine learning at {Facebook}:
  Understanding inference at the edge,'' in {\em HPCA}, 2019.

\bibitem{DNN_Compression_CharacterizingDNN_Mobile_Inference_TianGuo_arXiv_2019}
S.~S. Ogden and T.~Guo, ``Characterizing the deep neural networks inference
  performance of mobile applications,'' in {\em arXiv}, 2019,
  \url{https://arxiv.org/abs/1909.04783}.

\bibitem{DNN_Compression_PatDNN_Mobile_YanzhiWang_ASPLOS_2020}
W.~Liu, X.~Ma, S.~Lin, S.~Wang, X.~Qian, X.~Lin, Y.~Wang, and B.~Ren, ``Patdnn:
  Achieving real-time {DNN} execution on mobile devices with pattern-based
  weight pruning,'' in {\em ASPLOS}, 2020.

\bibitem{DNN_Compression_SongHan_ICLR_2016}
S.~Han, H.~Mao, and W.~J. Dally, ``Deep compression: Compressing deep neural
  networks with pruning, trained quantization and huffman coding,'' in {\em
  ICLR}, 2016.

\bibitem{DNN_EdgeInference_IntermittentEmbeddedSystem_CMU_ASPLOS_2019_Gobieski:2019:IBE:3297858.3304011}
G.~Gobieski, B.~Lucia, and N.~Beckmann, ``Intelligence beyond the edge:
  Inference on intermittent embedded systems,'' in {\em ASPLOS}, 2019.

\bibitem{DNN_Compression_Structured_YiranChen_NIPS_2016_10.5555/3157096.3157329}
W.~Wen, C.~Wu, Y.~Wang, Y.~Chen, and H.~Li, ``Learning structured sparsity in
  deep neural networks,'' in {\em NIPS}, 2016.

\bibitem{DNN_FBNet_HardwareAwareConvNetDesign_CVPR_2019_Wu2018FBNetHE}
B.~Wu, X.~Dai, P.~Zhang, Y.~Wang, F.~Sun, Y.~Wu, Y.~Tian, P.~Vajda, Y.~Jia, and
  K.~Keutzer, ``{FBNet}: Hardware-aware efficient {ConvNet} design via
  differentiable neural architecture search,'' in {\em CVPR}, pp.~10726--10734,
  2019.

\bibitem{DNN_Compression_AutoCompress_YanzhiWang_AAAI_2020}
N.~Liu, X.~Ma, Z.~Xu, Y.~Wang, J.~Tang, and J.~Ye, ``{AutoCompress}: An
  automatic dnn structured pruning framework for ultra-high compression
  rates,'' in {\em AAAI}, 2020.

\bibitem{DNN_Compression_PCONV_Sparsity_YanzhiWang_AAAI_2020}
X.~Ma, F.-M. Guo, W.~Niu, X.~Lin, J.~Tang, K.~Ma, B.~Ren, and Y.~Wang, ``Pconv:
  The missing but desirable sparsity in {DNN} weight pruning for real-time
  execution on mobile device,'' in {\em AAAI}, 2020.

\bibitem{DNN_EdgeInference_YiyuShi_NatureEle_2018_xu2018scaling}
X.~Xu, Y.~Ding, S.~X. Hu, M.~Niemier, J.~Cong, Y.~Hu, and Y.~Shi, ``Scaling for
  edge inference of deep neural networks,'' {\em Nature Electronics}, vol.~1,
  no.~4, p.~216, 2018.

\bibitem{DNN_ModelCompressionAcc_Survey_IEEE_TSP_2018_8253600}
Y.~{Cheng}, D.~{Wang}, P.~{Zhou}, and T.~{Zhang}, ``Model compression and
  acceleration for deep neural networks: The principles, progress, and
  challenges,'' {\em IEEE Signal Processing Magazine}, vol.~35, pp.~126--136,
  Jan 2018.

\bibitem{DNN_AugmentReality_JiasiChen_SenSys_2019_10.1145/3356250.3360044}
K.~Apicharttrisorn, X.~Ran, J.~Chen, S.~V. Krishnamurthy, and A.~K.
  Roy-Chowdhury, ``Frugal following: Power thrifty object detection and
  tracking for mobile augmented reality,'' in {\em SenSys}, 2019.

\bibitem{DNN_TensorFlowLite}
{TensorFlow}, ``Deploy machine learning models on mobile and {IoT} devices,''
  \url{https://www.tensorflow.org/lite/}.

\bibitem{han2015deep}
S.~Han, H.~Mao, and W.~J. Dally, ``Deep compression: Compressing deep neural
  networks with pruning, trained quantization and huffman coding,'' {\em arXiv
  preprint arXiv:1510.00149}, 2015.

\bibitem{han2015learning}
S.~Han, J.~Pool, J.~Tran, and W.~Dally, ``Learning both weights and connections
  for efficient neural network,'' in {\em Advances in neural information
  processing systems}, pp.~1135--1143, 2015.

\bibitem{lecun1990optimal}
Y.~LeCun, J.~S. Denker, and S.~A. Solla, ``Optimal brain damage,'' in {\em
  Advances in neural information processing systems}, pp.~598--605, 1990.

\bibitem{li2016pruning}
H.~Li, A.~Kadav, I.~Durdanovic, H.~Samet, and H.~P. Graf, ``Pruning filters for
  efficient convnets,'' {\em arXiv preprint arXiv:1608.08710}, 2016.

\bibitem{courbariaux2016binarized}
M.~Courbariaux, I.~Hubara, D.~Soudry, R.~El-Yaniv, and Y.~Bengio, ``Binarized
  neural networks: Training deep neural networks with weights and activations
  constrained to+ 1 or-1,'' {\em arXiv preprint arXiv:1602.02830}, 2016.

\bibitem{rastegari2016xnor}
M.~Rastegari, V.~Ordonez, J.~Redmon, and A.~Farhadi, ``Xnor-net: Imagenet
  classification using binary convolutional neural networks,'' in {\em European
  Conference on Computer Vision}, pp.~525--542, Springer, 2016.

\bibitem{denton2014exploiting}
E.~L. Denton, W.~Zaremba, J.~Bruna, Y.~LeCun, and R.~Fergus, ``Exploiting
  linear structure within convolutional networks for efficient evaluation,'' in
  {\em Advances in neural information processing systems}, pp.~1269--1277,
  2014.

\bibitem{lebedev2014speeding}
V.~Lebedev, Y.~Ganin, M.~Rakhuba, I.~Oseledets, and V.~Lempitsky, ``Speeding-up
  convolutional neural networks using fine-tuned cp-decomposition,'' {\em arXiv
  preprint arXiv:1412.6553}, 2014.

\bibitem{hinton2015distilling}
G.~Hinton, O.~Vinyals, and J.~Dean, ``Distilling the knowledge in a neural
  network,'' {\em arXiv preprint arXiv:1503.02531}, 2015.

\bibitem{srinivas2015data}
S.~Srinivas and R.~V. Babu, ``Data-free parameter pruning for deep neural
  networks,'' {\em arXiv preprint arXiv:1507.06149}, 2015.

\bibitem{molchanov2017variational}
D.~Molchanov, A.~Ashukha, and D.~Vetrov, ``Variational dropout sparsifies deep
  neural networks,'' in {\em Proceedings of the 34th International Conference
  on Machine Learning-Volume 70}, pp.~2498--2507, JMLR. org, 2017.

\bibitem{louizos2017learning}
C.~Louizos, M.~Welling, and D.~P. Kingma, ``Learning sparse neural networks
  through $ l\_0 $ regularization,'' {\em arXiv preprint arXiv:1712.01312},
  2017.

\end{thebibliography}
}

%\theendnotes

\end{document}